\documentclass[twocolumn, twocolappendix]{aastex631}
\usepackage{booktabs}
\usepackage{multirow}
\usepackage{ulem}
\usepackage{amsmath}
\usepackage{array}

\shorttitle{All-sky Galactic Soft X-ray Emission}
\shortauthors{Ampuku et al.}

\graphicspath{{./}{figures/}}

\begin{document}

\title{Probing the All-sky Distribution of Soft X-ray Emission Associated with the Milky Way}

\correspondingauthor{Ikuyuki Mitsuishi}
\email{mitsuisi@u.phys.nagoya-u.ac.jp}

\author[0009-0005-3295-7215]{Kazuki Ampuku}
\affiliation{Graduate School of Science, Tokai National Higher Education and Research System, Nagoya University, Furo-cho, Chikusa-ku, Nagoya, Aichi, 464-8602, Japan}

\author[0000-0002-9901-233X]{Ikuyuki Mitsuishi}
\affiliation{Graduate School of Science, Tokai National Higher Education and Research System, Nagoya University, Furo-cho, Chikusa-ku, Nagoya, Aichi, 464-8602, Japan}

\author[0000-0002-3638-0637]{Philip Kaaret}
\affiliation{NASA/Marshall Space Flight Center, Huntsville, AL 35812, USA}

\author[0000-0003-2907-0902]{Kyoko Matsushita}
\affiliation{Faculty of Physics, Tokyo University of Science, 1-3 Kagurazaka, Shinjuku-ku, Tokyo, 162-8601, Japan}

\author[0000-0001-8055-7113]{Kotaro Fukushima}
\affiliation{Faculty of Physics, Tokyo University of Science, 1-3 Kagurazaka, Shinjuku-ku, Tokyo, 162-8601, Japan}

\author[0000-0002-6631-6628]{Lorella Angelini}
\affiliation{NASA/Goddard Space Flight Center, Greenbelt, MD 20771, USA}

\author[0000-0002-5716-3412]{Dimitra Koutroumpa}
\affiliation{LATMOS/IPSL, UVSQ Universit\'e Paris-Saclay, Sorbonne Universit\'e, CNRS, Guyancourt, France}

\author{K. D. Kuntz}
\affiliation{NASA/Goddard Space Flight Center, Greenbelt, MD 20771, USA}
\affiliation{The Henry A. Rowland Department of Physics and Astronomy, Johns Hopkins University, Baltimore, MD 21218, USA}

\begin{abstract}

We present an all-sky analysis of soft X-ray emission associated with the Milky Way using \textit{HaloSat} and \textit{ROSAT} data, extending our previous study of five Galactic-plane regions to 330 fields across the sky.
We performed uniform spectral fitting with a model including an unabsorbed Local Hot Bubble and two additional absorbed thermal components, representing emission from warm-hot and hot plasma, while accounting for solar wind charge exchange and the cosmic X-ray background.
The warm-hot component is often interpreted as emission from the circumgalactic medium, whereas the origin of the hot component remains uncertain.
Individual-field fits generally favored the inclusion of the hot component, and the field-to-field temperature distributions had medians of $kT=0.19$ and $0.74$~keV, with 16th--84th percentile ranges of $0.17$--$0.21$ and $0.61$--$0.90$~keV, respectively.
The corresponding 16th--84th percentile EM ranges were $(1.0$--$8.2)\times10^{-2}$ and $(0.5$--$4.1)\times10^{-3}~\mathrm{cm^{-6}\,pc}$.
Stacked spectra from fields within $60^\circ\leq l\leq180^\circ$ supported the individual-field results.
After excluding bright X-ray sources and large-scale Galactic structures, we modeled the EM distributions of 178 fields using a hybrid spatial framework comprising stellar emission and disk-like and halo-like diffuse-plasma components.
The warm-hot emission was described mainly by the disk-like term, whereas the hot emission was described mainly by stellar emission with an additional halo-like term.
These results support a widespread two-temperature description and suggest a composite origin for the hot emission.

\end{abstract}


\section{Introduction} \label{sec:intro}

Multiwavelength observations have revealed that galaxies are surrounded by multiphase gas spanning a wide temperature range of $10^4$--$10^7$ K \citep{Tumlinson+2017}.
The gas surrounding galaxies is commonly referred to as the circumgalactic medium (CGM), which forms an interface between the galactic disk, the interstellar medium (ISM), and the intergalactic medium.
The CGM plays a key role in the baryon cycle of galaxies through gas accretion, outflows from the galactic disk, heating and cooling, and metal transport \citep[e.g.,][]{Putman+2012, Chen+2026}.
Among the different phases of the CGM, the hot phase traced by X-ray emission is important for understanding galaxy evolution, because X-ray-emitting plasma can retain information on gas heated in the galactic gravitational potential and on past energy injection by stellar feedback and AGN activity.
The Milky Way provides the nearest laboratory for probing the spatial distribution of the hot phase from an internal vantage point.
However, observations from inside the Galactic disk inevitably include multiple X-ray emission components superposed along each line of sight.
Therefore, understanding the X-ray emission associated with the Milky Way requires adequate energy resolution for separating individual emission components, as well as broad sky coverage over many lines of sight.

Soft X-ray emission and absorption-line studies of the Milky Way have widely reported a thermal plasma component at $\sim$0.2~keV, hereafter referred to as the warm-hot component \citep[e.g.,][]{Yoshino+2009, Gatuzz+2017, Kaaret+2020, Mathur+2021}.
The warm-hot component is often interpreted as emission from the CGM or Galactic halo gas, at a temperature close to the virial temperature of the Milky Way.
Some studies have also explored models allowing enhanced N and/or Ne abundances for this component \citep{Gupta+2021, Gupta+2023, Gupta+2025}.
In recent years, an additional hotter, super-virial component, hereafter referred to as the hot component, has been reported along several lines of sight, with characteristic temperatures of 0.5--1~keV \citep[e.g.,][]{Das+2019a, Gupta+2021, Bhattacharyya+2023, Sugiyama+2023}.
The origin of this hot component has been discussed mainly in terms of two broad possibilities.
The first possibility is emission associated with stars, particularly coronal emission from low-mass stars.
For example, \citet{Masui+2009} and \citet{Wulf+2019} focused on dwarf M-type stars (dM stars) among low-mass stellar populations and showed that unresolved dM stars can explain the $\sim$0.8--0.9~keV emission observed at low Galactic latitudes.
\citet{Ampuku+2024} also detected hot X-ray emission from regions near the Galactic plane and suggested that part of the hot emission may originate from stellar sources.
Recent \textit{SRG}/eROSITA observations have also reported a correspondence between the hot component and the stellar mass profile of the Milky Way, and have shown that the average X-ray spectra of nearby low-mass stars contain two thermal components \citep{Knies+2024, Zheng+2026}.
The second possibility is a diffuse plasma component.
\citet{Das+2021} and \citet{Bhattacharyya+2023} proposed that outflows driven by star formation in the Galactic disk may produce a super-virial hot component that extends inhomogeneously into the halo.
\citet{Roy+2025} and \citet{Gupta+2025} further argued that the hot component may not be confined to the local ISM but may also be distributed outside the Galactic disk.
The origin of the hot component therefore remains uncertain and may have a single dominant origin or may instead represent a hybrid component in which stellar emission and diffuse plasma both contribute.

Previous studies with broad sky coverage have provided important constraints on the origin of the hot component.
\citet{Bluem+2022} used the wide-field observations of \textit{HaloSat} to measure the temperatures and emission measures (EMs) of the warm-hot and hot components at high Galactic latitudes.
The analysis by \citet{Bluem+2022} provided observational evidence for a two-temperature structure but was mainly focused on high Galactic latitudes, where the contribution from stellar emission is expected to be relatively small.
Therefore, the possible stellar contribution at low Galactic latitudes and the potential of the all-sky spatial distribution to constrain the origin of the emission were not fully examined.
\citet{Ponti+2026} used \textit{SRG}/eROSITA data from the western Galactic hemisphere to show a correspondence between the EM of the hot component and the stellar mass distribution, suggesting that low-mass stars provide a major contribution.
However, because \citet{Ponti+2026} evaluated the EM distribution with the temperature of the hot component fixed at 0.7~keV, the temperature distribution and its possible inhomogeneity were not examined.
If the temperature of the hot component varies among different lines of sight, the spatial behavior of the temperature may provide useful clues for distinguishing among local X-ray emission, stellar emission, and diffuse plasma.
Therefore, to investigate the origin of the hot component in more detail, it is important to measure the temperatures and EMs of both the warm-hot and hot components through a uniform all-sky spectral analysis and to compare stellar emission and diffuse plasma within the same spatial modeling framework.

In this study, we extend the analysis of \citet{Ampuku+2024}, which was limited to regions near the Galactic plane, and systematically investigate the nature of the hot component and its possible contributors using the all-sky \textit{HaloSat} data set together with \textit{ROSAT} R1/R2 constraints.
In particular, we uniformly analyze the all-sky distributions of the warm-hot and hot components, while accounting for solar wind charge exchange (SWCX) emission \citep[see][for a review]{Kuntz+2019}, which can affect the quantitative measurements of their temperatures and EMs.
We then apply a hybrid spatial framework incorporating stellar emission and diffuse plasma to constrain their relative contributions.

\section{Observations} \label{sec:obs}

\textit{HaloSat} \citep{Kaaret+2019} was a CubeSat designed to map soft X-ray emission over the entire sky that operated from 2018 to 2021. 
\textit{HaloSat} observations were conducted during orbital night, reducing the background in combination with shielding by the Earth's magnetosphere. 
The three silicon drift detectors onboard \textit{HaloSat} covered the 0.4--7~keV energy band, had sufficient energy resolution to distinguish emission lines such as O\,\textsc{vii} and O\,\textsc{viii}, and provided a wide field of view of $\sim$100~deg$^2$ \citep{Zajczyk+2020}.
Because \textit{HaloSat} was a non-imaging instrument, contaminating sources within the field of view could not be removed spatially and had to be assessed separately.
Nevertheless, the wide field of view provided \textit{HaloSat} with a large grasp, the product of effective area and field-of-view solid angle. This characteristic made \textit{HaloSat} a powerful instrument for observing faint diffuse X-ray emission, including emission from the CGM, with high signal-to-noise ratios \citep[e.g.,][]{Silich+2021, Ringuette+2021, Bluem+2022, Ampuku+2024}.

In this study, we used data from the \textit{HaloSat} archive at the HEASARC\footnote{\url{https://heasarc.gsfc.nasa.gov/FTP/halosat/data/obs/}\\(Ver. 20230501)}.
Among the 372 all-sky observation fields, we included all 344 fields for which standard screening had been applied and the mean exposure time of the three detectors was longer than 5 ks.
We used the latest version of the \textit{HaloSat} calibration database\footnote{\url{https://heasarc.gsfc.nasa.gov/docs/halosat/caldb/index.html}\\(Ver. 20230324)} for the detector responses.
We also used the all-sky diffuse background maps from \textit{ROSAT} \citep{Snowden+1997}.
The R1 (0.11--0.284~keV) and R2 (0.14--0.284~keV) bands are useful for constraining low-temperature foreground components such as the Local Hot Bubble (LHB).
Following \citet{Yeung+2024}, we extracted the R1 and R2 count rates and uncertainties corresponding to each \textit{HaloSat} observation field and included them as two additional constraints in the spectral analysis.

Spectral analysis was performed using HEAsoft v6.33.2 and PyXspec version 2.1.2 in XSPEC version 12.13.1e \citep{Arnaud+1996, Gordon+2021}.
For the plasma emission models, we used the Astrophysical Plasma Emission Code \citep[\texttt{apec} and \texttt{vapec} models in XSPEC;][]{Smith+2001}, assuming optically thin plasma in collisional ionization equilibrium.
We adopted the abundance table of \citet{Wilms+2000} and the photoelectric absorption cross sections of \citet{Verner+1996}.
Motivated by the treatment of \citet{Yeung+2024}, we defined the total fit statistic as the sum of the C statistic \citep{Cash+1979} for the \textit{HaloSat} spectra and the $\chi^2$ statistic for the \textit{ROSAT} R1/R2 count rates.
We explored the posterior distributions of the model parameters using Markov Chain Monte Carlo (MCMC) sampling with the Goodman--Weare algorithm.
Using previous studies \citep[e.g.,][]{Yeung+2023} as a guide, we used 40 walkers with 10,000 steps for each fit and discarded the first 2,500 steps as burn-in.
Throughout this paper, parameter values are reported as posterior medians, with uncertainties corresponding to the 16th--84th percentile ranges.

\section{Analysis and Results} \label{sec:analysis_results}

\subsection{Spectral Fitting Method}

Following \citet{Bluem+2022} and \citet{Ampuku+2024}, we adopted a six-component spectral model consisting of SWCX and the LHB as unabsorbed foreground emission, the cosmic X-ray background (CXB) and the warm-hot and hot components subject to interstellar absorption, and the instrumental background.
Because \textit{HaloSat} and \textit{ROSAT} were observed at different epochs, we treated their SWCX contributions independently.
For \textit{HaloSat} data, following previous studies \citep{Koutroumpa+2012, Kaaret+2020}, we estimated the line-of-sight-integrated SWCX line intensities corresponding to each observation 
using heliospheric simulations with highly charged ion abundance input data from the \textit{ACE} satellite \citep{Gloeckler+1998}.
In addition to O\,\textsc{vii} and O\,\textsc{viii}, which were considered in \citet{Kaaret+2020} and \citet{Bluem+2022}, we included C\,\textsc{v}, C\,\textsc{vi}, Ne\,\textsc{ix}, and Mg\,\textsc{xi} lines that may affect the quantitative measurements of the warm-hot and hot components.
These SWCX lines were modeled with \texttt{gaussian} components, with their line energies and intensities fixed.
For \textit{ROSAT} data, we used the SWCX contribution models of \citet{Uprety+2016} and \citet{Liu+2017}, averaged over each observed field, as fixed contributions to the R1 and R2 bands.
The LHB was modeled with an unabsorbed \texttt{apec} component. Its abundance and temperature were fixed at $Z=1.0\,Z_{\odot}$ and $kT=0.1$~keV, respectively \citep{Liu+2017}, and only the emission measure (EM) was left free.
The CXB was represented by an absorbed \texttt{powerlaw} model, with the photon index fixed at $\Gamma = 1.412$ and the 2--10~keV flux constrained to $(6.4\pm0.6)\times10^{-8}~ \mathrm{erg~cm^{-2}~s^{-1}~sr^{-1}}$ \citep{Kushino+2002}.
The warm-hot and hot components were described by absorbed \texttt{apec} models with abundances fixed at $Z=0.3\,Z_{\odot}$ and $Z=1.0\,Z_{\odot}$, respectively \citep{Ampuku+2024}. Their temperatures and EMs were treated as free parameters.
Interstellar absorption was modeled with \texttt{tbabs}.
We adopted the field-averaged $N_{\rm H}$ values calculated from the \textit{Planck} thermal dust map \citep{Planck+2014, Zhu+2017}.
The absorption column densities for the CXB and the warm-hot component were fixed to these calculated values.
For the hot component, whose line-of-sight extent is uncertain, the absorption column density was treated as a free parameter with the calculated value as an upper limit.
The instrumental background was modeled following the latest \textit{HaloSat} calibration documents\footnote{\url{https://heasarc.gsfc.nasa.gov/docs/halosat/analysis/halosat_analysis_20221205.pdf}, \url{https://heasarc.gsfc.nasa.gov/docs/halosat/analysis/back20221130.pdf}}, using \texttt{powerlaw} components with the photon indices of the soft and hard components fixed.

\begin{figure}[htbp]
\centering
\includegraphics[width=\linewidth]{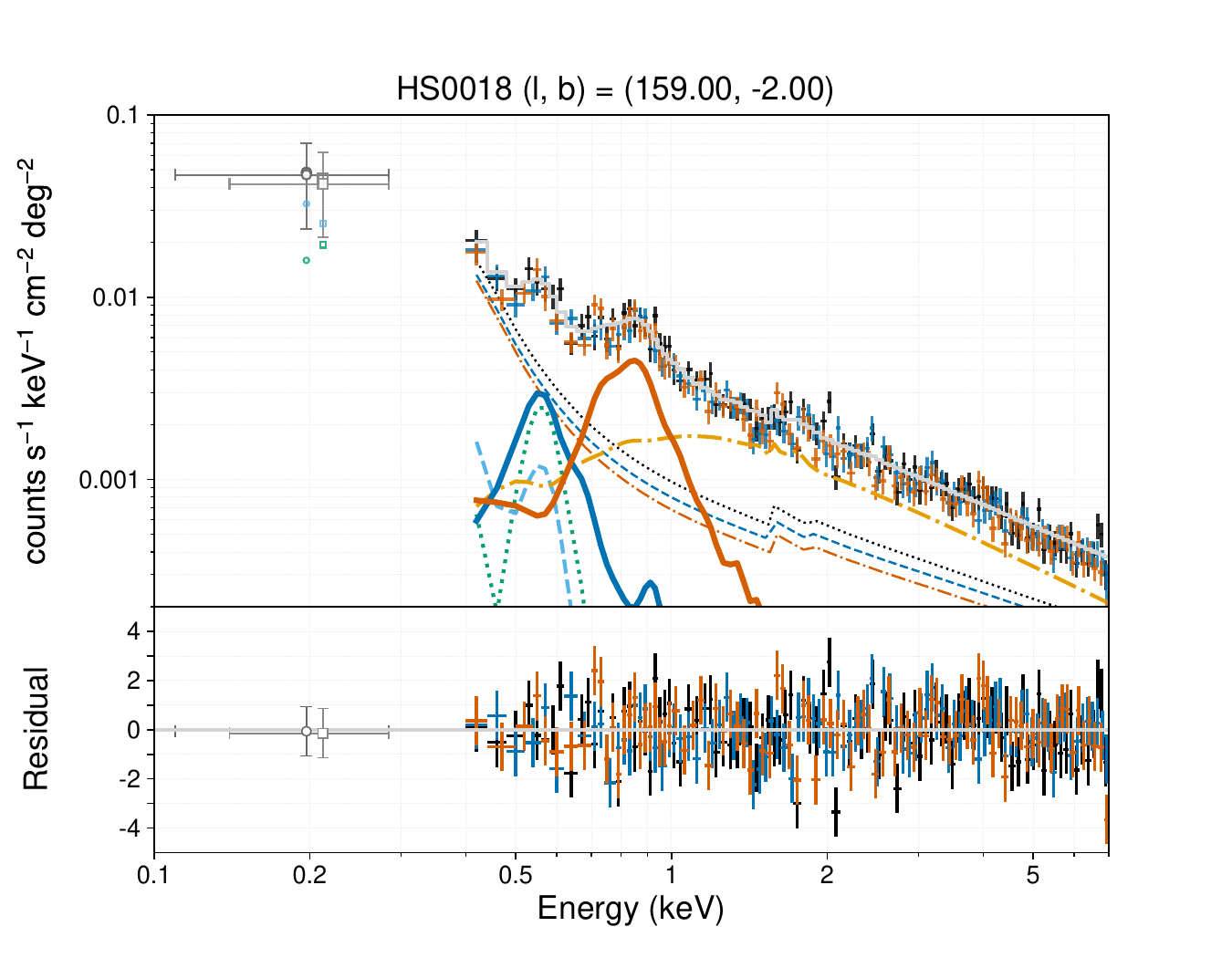}
\vspace{-1.0em}
\includegraphics[width=\linewidth]{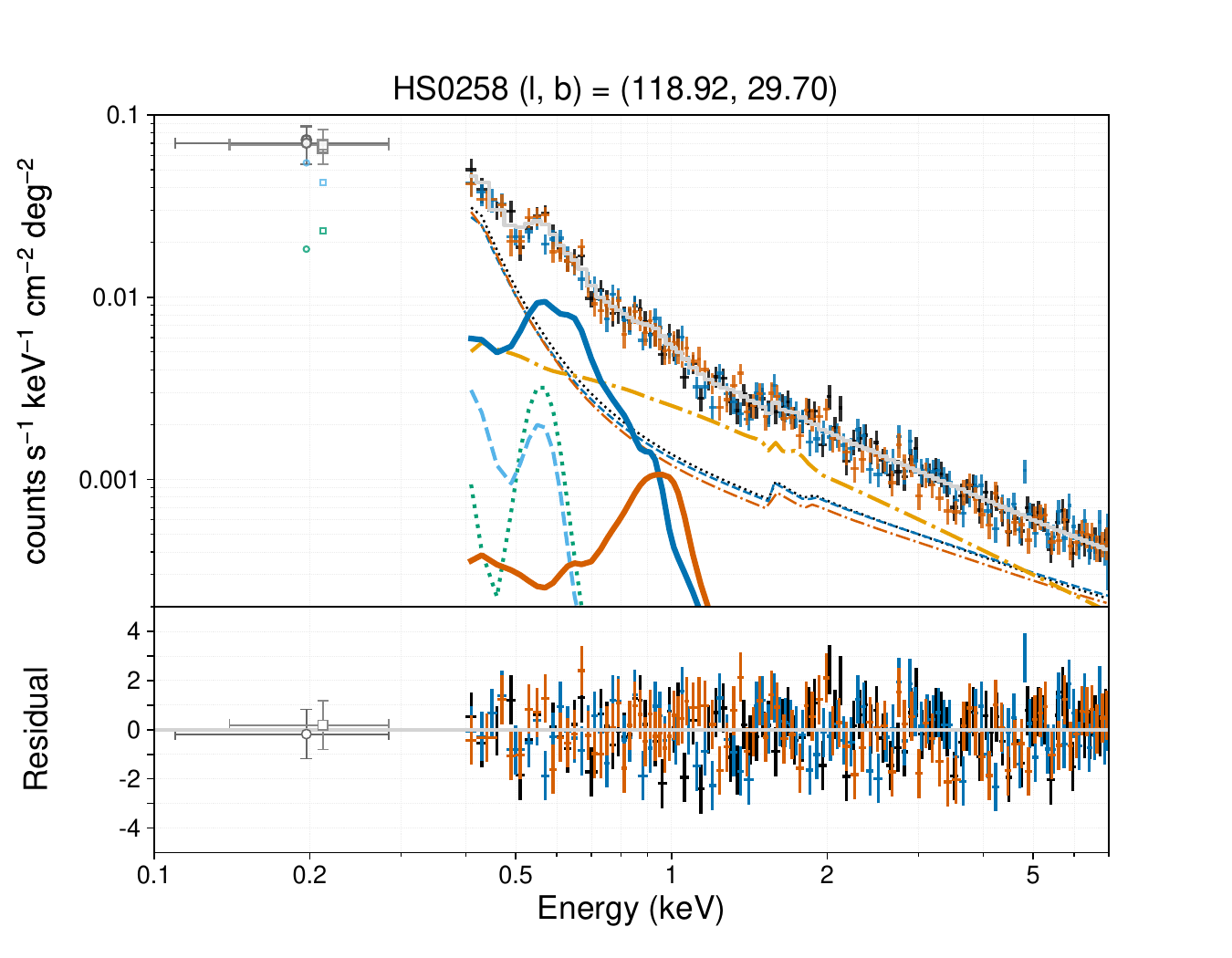}
\caption{Examples of the spectral fitting results for representative fields. The upper panels show the observed spectra and fitted models. The black, blue, and vermilion crosses represent the data from the three \textit{HaloSat} detectors, while the gray open circles and squares show the two \textit{ROSAT} data points.
The light-gray curves show the total models. The individual model components are shown as SWCX (green dotted), LHB (sky-blue dashed), CXB (orange dash-dotted), the warm-hot component (blue solid), and the hot component (vermilion solid). The instrumental backgrounds for the three detectors are shown as black dotted, blue dashed, and vermilion dash-dotted lines. The lower panels show the residuals in units of the statistical uncertainties. The spectra are binned for display purposes only.}
\label{fig:example_spec}
\end{figure}

Following the \textit{HaloSat} data-fitting strategy of \citet{Ampuku+2024}, we performed simultaneous fits by linking the astrophysical emission components among the three \textit{HaloSat} detectors, while treating the instrumental background independently for each detector.
We also included the two \textit{ROSAT} R1/R2 data points in the same simultaneous fit using the same astrophysical emission model apart from the independently treated SWCX contribution.
The spectral fits were performed on unbinned data.
As an auxiliary measure of goodness of fit, we grouped the \textit{HaloSat} spectra to have at least 25 counts per bin and calculated $\chi^2$/DoF using the same spectral model \citep[e.g.,][]{Yeung+2024}.
We found that all fields with $\chi^2$/DoF $\geq$ 1.5 contained bright X-ray sources listed in the MAXI Solid-state Slit Camera all-sky catalog \citep{Tomida+2016} within the \textit{HaloSat} field of view.
Hereafter, these sources are referred to as MAXI sources.
Because emission from these sources cannot be spatially separated in the \textit{HaloSat} data and can bias the spectral decomposition, we excluded these fields from the quantitative analysis below.
The final sample consisted of 330 fields.
Examples of the fitting results are shown in Figure~\ref{fig:example_spec}.

In the MCMC-based uncertainty estimation, the posterior distributions of the thermal-component temperatures did not show a unique peak in some fields, indicating degeneracy among the LHB, warm-hot component, and hot component \citep[cf.][]{Yeung+2024}.
This uncertainty is likely due to the limited photon statistics and the difficulty in constraining the plasma temperature from the spectral structure.
We systematically identified poorly constrained temperatures using binned histograms of the posterior samples, classifying a component as poorly constrained when the histogram showed multiple comparable peaks, lacked an isolated dominant peak, or was concentrated near a parameter boundary.
For such fields, following the approach adopted in previous work \citep{Bluem+2022}, we refitted the spectra after fixing the temperature of the poorly constrained component to its representative value, either 0.2 or 0.7~keV.

\subsection{Individual-field Results} \label{sec:results}

Observation information and spectral-fitting results for the final sample of 330 fields are listed in Table~\ref{tab:best-fit}.
Unless otherwise noted, all ranges quoted below correspond to the 16th--84th percentiles of the field-to-field distributions in this sample rather than uncertainties from individual spectral fits.

We first compared a single-temperature model containing only the warm-hot component with a two-temperature model including both the warm-hot and hot components.
We evaluated the improvement using the \texttt{ftest} command in XSPEC as an empirical indicator rather than a strict hypothesis test.
The two-temperature model improved the fits at levels corresponding to $>2\sigma$ in 75\% (249/330) of the fields and $>3\sigma$ in 60\% (198/330).
A comparison using the XSPEC \texttt{goodness} command showed the same overall trend.
We therefore adopted the two-temperature fitting results to characterize the two components below.

The SWCX contribution was characterized mainly by the O\,\textsc{vii} line.
Its contribution to the 0.5--0.65~keV flux ranged from 4\% to 8\%, reaching up to $\sim$30\% in some fields.
The O\,\textsc{vii} and carbon-line intensities were comparable, whereas the O\,\textsc{viii} and Ne\,\textsc{ix} intensities were approximately one order of magnitude lower and the Mg\,\textsc{xi} intensity was another order of magnitude lower.
The LHB EMs ranged from $(1.5$--$5.3) \times 10^{-3}~\mathrm{cm^{-6}\,pc}$.
The calculated $N_{\rm H}$ values ranged from $(0.02$--$0.26) \times 10^{22}~\mathrm{cm^{-2}}$.
In contrast, the fitted $N_{\rm H}$ values for the hot component ranged from $(0.01$--$0.15) \times 10^{22}~\mathrm{cm^{-2}}$ and tended to be lower than the calculated values.
The temperature of the warm-hot component was fixed in eight fields, and that of the hot component was fixed in 102 fields.
These fields were excluded when calculating the field-to-field temperature distributions.
The remaining fields yielded median temperatures of $kT = 0.19$ and $0.74$~keV, with 16th--84th percentile ranges of $0.17$--$0.21$ and $0.61$--$0.90$~keV, respectively.
The EMs of the warm-hot and hot components had corresponding ranges of $(1.0$--$8.2) \times 10^{-2}$ and $(0.5$--$4.1) \times 10^{-3}~\mathrm{cm^{-6}\,pc}$, respectively.
We also confirmed that the normalizations of the two instrumental-background components were within their typical ranges for all three detectors.

\begin{figure*}[htbp]
    \begin{center}
    \includegraphics[width=1.0\linewidth]{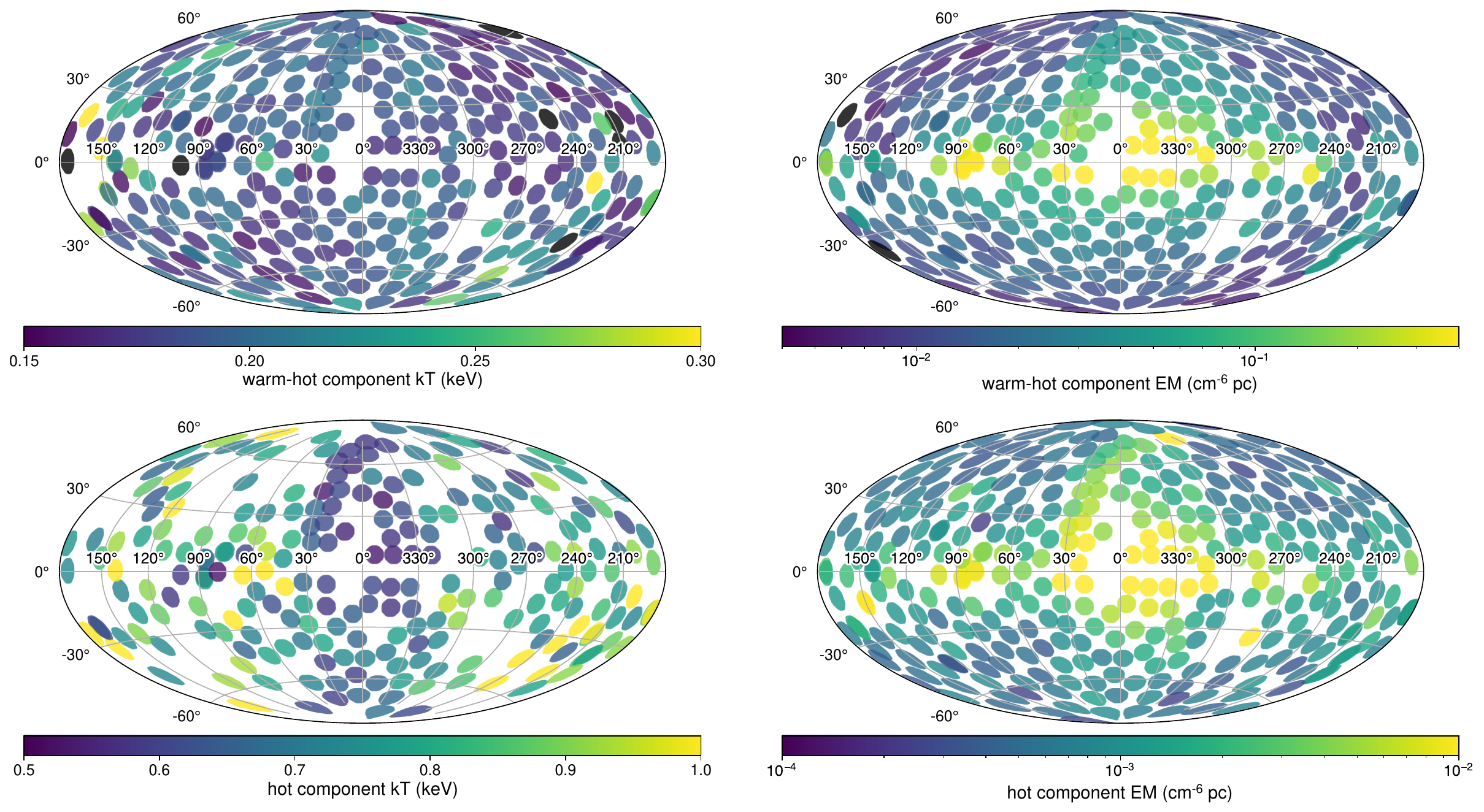}
    \end{center}
    \caption{Fitting results for the all-sky observation fields shown in a Hammer--Aitoff projection in Galactic coordinates. The upper-left and upper-right panels show the $kT$ and EM of the warm-hot component, respectively, while the lower-left and lower-right panels show those of the hot component. Each observation field is displayed as a circle with a radius of 6\textdegree. Fields in which the temperature of each component was fixed are not included in the corresponding $kT$ panels.}
    \label{fig:all-sky_results}
\end{figure*}

Figure~\ref{fig:all-sky_results} shows the all-sky spatial distributions of the warm-hot and hot components.
The temperature of the warm-hot component is generally distributed around $\sim$0.2~keV, although local deviations are present in some fields.
Compared with the warm-hot component, the hot component shows clearer spatial variations in temperature, with lower values of $\sim$0.5~keV toward the eROSITA bubbles and higher values of $\gtrsim$1~keV along several other lines of sight.
The EMs of both components show broadly similar spatial dependence and are enhanced toward the Cygnus and Orion--Eridanus superbubbles, the eROSITA bubbles, and the Galactic plane.
Toward the eROSITA-bubble region and outside this region, the warm-hot component had 16th--84th percentile temperature ranges of $0.18$--$0.21$ and $0.17$--$0.21$~keV and EM ranges of $(2$--$11) \times 10^{-2}$ and $(0.9$--$2.7) \times 10^{-2}~\mathrm{cm^{-6}\,pc}$, respectively.
The corresponding ranges for the hot component were $0.59$--$0.77$ and $0.67$--$0.92$~keV in temperature and $(0.9$--$4.7) \times 10^{-3}$ and $(0.4$--$1.1) \times 10^{-3}~\mathrm{cm^{-6}\,pc}$ in EM.

\begin{figure}[htbp]
    \centering
    \includegraphics[width=\linewidth]{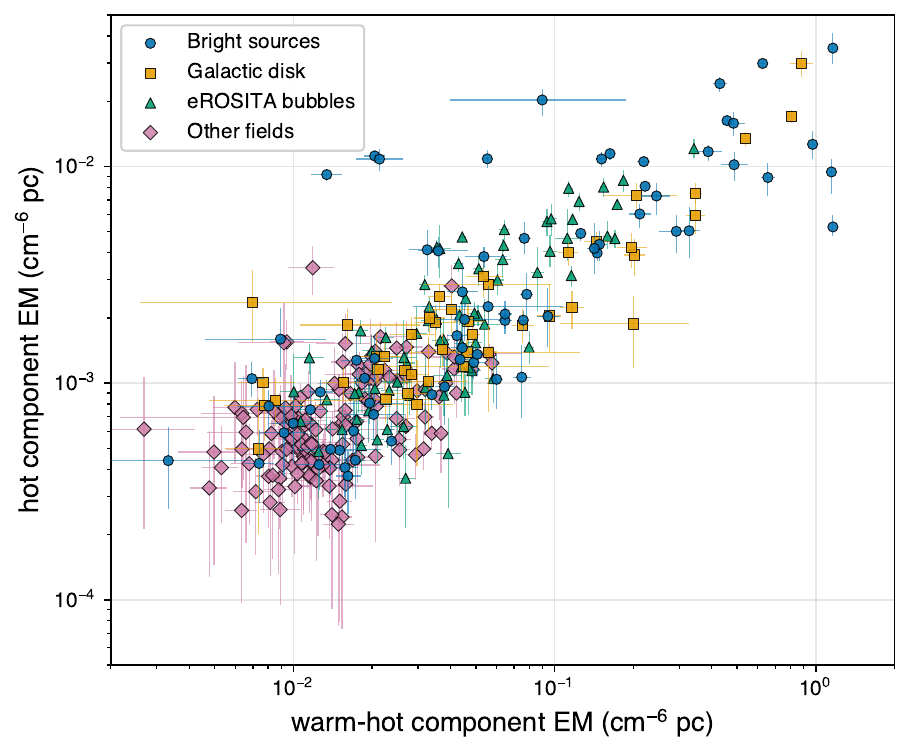}
    \caption{
    Relationship between the EMs of the warm-hot and hot components for the 330 individual fields.
    The fields are assigned, in priority order, to bright-source regions containing the Cygnus and Orion--Eridanus superbubbles or MAXI sources (blue circles), Galactic-disk regions with $|b|\leq15^\circ$ (orange squares), fields toward the eROSITA bubbles with $|l|\leq60^\circ$ (green triangles), or all other fields (purple diamonds).
    }
    \label{fig:corr_em}
\end{figure}

Using the individual-field fitting results, we examined the correlations among the $N_{\rm H}$, $kT$, and EM values of the warm-hot and hot components.
The full set of parameter distributions and pairwise relationships is presented in Appendix Figure~\ref{fig:corner_region}.
For all 330 fields, the EMs of the two components show a positive Spearman rank correlation of $\rho=0.75^{+0.02}_{-0.02}$.
To examine the influence of spatial structures, we classified the fields into four mutually exclusive categories.
The fields were assigned, in priority order, to bright-source regions containing the Cygnus and Orion--Eridanus superbubbles or MAXI sources, Galactic-disk regions with $|b|\leq15^\circ$, fields toward the eROSITA bubbles with $|l|\leq60^\circ$, and all other fields.
As shown in Figure~\ref{fig:corr_em}, fields in the bright-source, Galactic-disk, and eROSITA-bubble categories preferentially occupy the higher-EM ranges.
Fields in the Other category span approximately one order of magnitude in each EM and show a weaker correlation of $\rho=0.43^{+0.05}_{-0.05}$.
The quoted coefficients are the medians of distributions obtained by sampling the EMs from their MCMC posteriors, with uncertainties corresponding to the 16th--84th percentiles.

We next compared these results with previous studies.
The LHB EMs obtained in this work are typically 0.5--2 times those reported by \citet{Liu+2017}, indicating broad agreement with some field-to-field differences.
While \citet{Liu+2017} inferred the LHB properties primarily from the \textit{ROSAT} R2/R1 band ratio, we determined the LHB EM simultaneously with the other emission components through joint spectral modeling of the \textit{HaloSat} and \textit{ROSAT} data.
In particular, explicitly including both the warm-hot and hot components may improve the separation of the LHB emission from more distant thermal emission and may therefore partly account for the field-to-field differences between the two analyses.
The resulting all-sky LHB EM distribution, shown in Appendix Figure~\ref{fig:lhb_appendix}, provides an independent estimate based on spectral-component decomposition.
For the warm-hot and hot components, the temperatures are consistent with those reported in recent observational studies assuming a two-temperature structure \citep[e.g.,][]{Gupta+2021, Ueda+2022}.
Some differences in EM can be largely attributed to the different assumed metal abundances, because the EM is approximately inversely proportional to the assumed abundance for spectra obtained with CCD-like energy resolution.\footnote{If the abundances are denoted by $Z$ and $Z'$, the corresponding EMs can be approximated as EM$_{Z'} \simeq (Z/Z')$ EM$_Z$ \citep[e.g.,][]{Ueda+2022, Bluem+2022}. We confirmed that the EM follows this scaling in our spectral model.}

We also compared our results with previous \textit{HaloSat} studies.
For the five fields shared with \citet{Ampuku+2024}, the hot-component temperatures reported by \citet{Ampuku+2024} are consistent with those obtained in the present work, although our EMs are up to a factor of $\sim$2 larger, likely because absorption is included for the hot component.
For the fields shared with \citet{Bluem+2022}, the temperatures are generally consistent within the uncertainties, while our EMs are systematically higher by roughly 50\% after abundance scaling, likely reflecting differences in the instrumental background modeling.
Because the following analysis focuses on relative spatial variations derived with a uniform model, we consistently use the parameters obtained in the present work.

Despite the different angular scales and sight-line sampling, the temperature ranges obtained here are consistent with those reported by \citet{Gupta+2023}, who found warm-hot temperatures of approximately 0.20~keV and hot-component temperatures spanning approximately 0.4--1.2~keV. In both analyses, the temperature distributions of both components overlap inside and outside the eROSITA bubbles, whereas their EMs are enhanced inside the bubbles.
Compared with \citet{Ponti+2026}, which analyzed a half-sky data set, the hot EM distribution in this work shows a similar two-order-of-magnitude inhomogeneity.
A direct comparison with the latitude profile shown in their Figure~3 confirms that the hot EMs are consistent within the uncertainties over the longitude range $220^\circ < l < 235^\circ$.

Finally, previous two-temperature studies reported positive correlations between the EMs of the warm-hot and hot components, suggesting a possible physical connection between them \citep[e.g.,][]{Bluem+2022, Bhattacharyya+2023}.
We also find a positive all-sky correlation, but it becomes substantially weaker after excluding high-EM fields associated with specific Galactic structures.
This weakening indicates that the all-sky correlation is partly driven by enhancements of both components within the same structures and therefore does not necessarily imply a common physical origin.
This result motivates further spatial modeling of the EM distributions.

\subsection{Stacked-field Analysis}

To improve photon statistics and examine the individual-field results on broader spatial scales, we performed a stacked spectral analysis by combining spectra from multiple fields.
We selected the Galactic longitude range $60^\circ\leq l\leq180^\circ$ to minimize the effects of the eROSITA bubbles and other large-scale Galactic structures, such as the Orion--Eridanus superbubble and the Antlia supernova remnant.
Within this longitude range, fields overlapping the Cygnus superbubble or containing MAXI sources were excluded.
The remaining fields were grouped into eight stacked regions, A--H, according to Galactic latitude (Table~\ref{table:best-fit-stack}).
We used the same spectral model as in the individual-field analysis.
For the \textit{ROSAT} data, $N_{\rm H}$, and SWCX line intensities, we adopted exposure-weighted averages of the values for the fields included in each stacked region.
Because a discontinuous structure associated with data screening becomes apparent around 3~keV in stacked spectra \citep{Bluem+2022}, we excluded the 2.4--3.0~keV band.

\begin{figure}[tbhp]
\centering
\includegraphics[width=\linewidth]{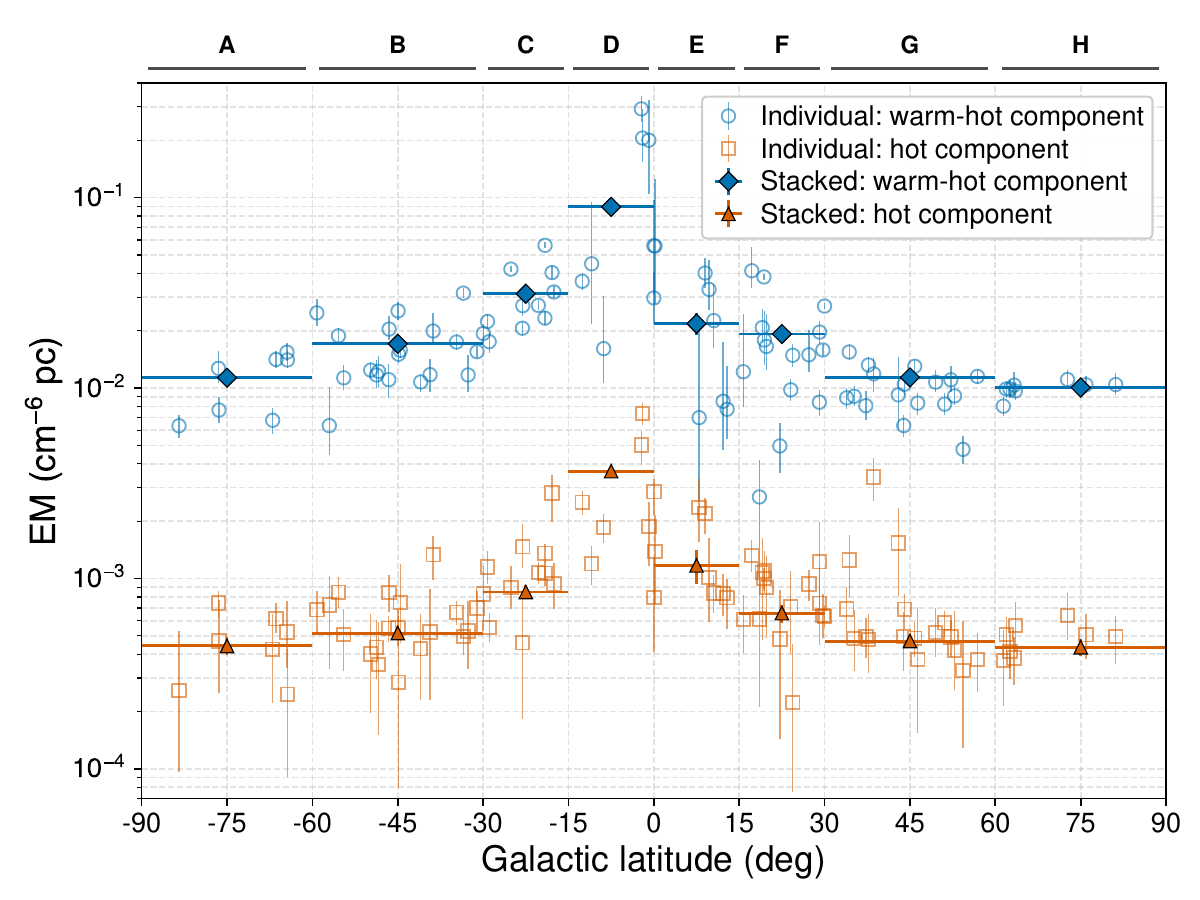}
\caption{Distributions of the EMs of the individual and stacked regions as a function of Galactic latitude. Open blue circles and open vermilion squares represent the warm-hot and hot components in individual fields, respectively. Filled blue diamonds and filled vermilion triangles represent the warm-hot and hot components in the stacked regions, respectively. The horizontal bars labeled A--H at the top denote the eight stacked Galactic-latitude intervals listed in Table~\ref{table:best-fit-stack}.}
\label{fig:stack}
\end{figure}

Using the same empirical model-comparison procedure as for the individual fields, all eight stacked regions favored the two-temperature model over the single-temperature model at levels corresponding to $>3\sigma$.
The fitted temperatures ranged from 0.16 to 0.19~keV for the warm-hot component and from 0.64 to 0.81~keV for the hot component (Table~\ref{table:best-fit-stack}).
The EMs of both components were approximately eight times higher at low than at high Galactic latitudes.
The stacked results therefore support the temperature and latitude-dependent EM trends obtained from the individual-field analysis with improved photon statistics (Figure~\ref{fig:stack}).

For Stack H, which shares many individual fields with the stacked region analyzed by \citet{Bluem+2022}, we confirmed that the temperatures and EMs obtained in the present work are comparable to those reported by \citet{Bluem+2022}.

We next examined whether abundance variations in the warm-hot component could provide an alternative to the standard two-temperature model.
To test the single-temperature model with enhanced Ne considered by \citet{Gupta+2021}, we replaced the \texttt{apec} model of the warm-hot component with \texttt{vapec} and allowed the Ne abundance to vary, with all other abundances fixed at solar values.
The standard two-temperature model was preferred over this Ne-variable single-temperature model at levels corresponding to $>3\sigma$ in all eight regions.

We also tested the two-temperature model with enhanced N and Ne considered by \citet{Gupta+2025}.
The warm-hot component was described with \texttt{vapec}, with its N and Ne abundances treated as free parameters while the other elemental abundances were fixed to solar values, whereas the hot component was described with \texttt{apec}.
This model yielded improvements above $3\sigma$ over the standard two-temperature model in Stacks B, G, and H.
Both N and Ne were enhanced only in Stack H, whereas Stacks B and G showed enhanced N alone.
Thus, on the broad spatial scales examined here, models including both N and Ne enhancements did not generally provide a statistically preferred alternative to the standard two-temperature model.

Because these model comparisons can depend on the adopted $N_{\rm H}$ values, we repeated them using alternative absorption-column maps \citep[e.g.,][]{Rachford+2002, Willingale+2013, HI4PI+2016}.
Although the significance of individual comparisons changed, the main conclusions were unchanged.

\begin{deluxetable*}{ccccccccccc}
    \tablecaption{Fitting results for the stacked spectra with the two-temperature model.}\label{table:best-fit-stack}
    \tabletypesize{\footnotesize}
    \tablehead{ Stack & $b$ & \multicolumn{3}{c}{warm-hot component} & & \multicolumn{3}{c}{hot component} & & $\chi^2$/DoF \\\cline{3-5} \cline{7-9} & (deg.) & $N_{\textrm{H}}$\tablenotemark{a} & $kT$\tablenotemark{b} & EM\tablenotemark{c} & & $N_{\textrm{H}}$\tablenotemark{a} & $kT$\tablenotemark{b} & EM\tablenotemark{d} & & }
    \startdata
    A & $-90$ to $-60$ & $0.027 \textrm{ (fixed)}$ & $0.18^{+0.01}_{-0.01}$ & $1.1^{+0.1}_{-0.1}$ &  & $\le 0.027$ & $0.64^{+0.05}_{-0.06}$ & $0.44^{+0.05}_{-0.05}$ &  & $925 / 883$ \\
    B & $-60$ to $-30$ & $0.062 \textrm{ (fixed)}$ & $0.18^{+0.01}_{-0.01}$ & $1.7^{+0.1}_{-0.1}$ &  & $0.034^{+0.028}_{-0.022}$ & $0.71^{+0.03}_{-0.03}$ & $0.52^{+0.04}_{-0.04}$ &  & $977 / 883$ \\
    C & $-30$ to $-15$ & $0.097 \textrm{ (fixed)}$ & $0.19^{+0.01}_{-0.01}$ & $3.1^{+0.1}_{-0.1}$ &  & $<0.073$ & $0.77^{+0.02}_{-0.02}$ & $0.85^{+0.08}_{-0.06}$ &  & $890 / 883$ \\
    D & $-15$ to $0$ & $0.39 \textrm{ (fixed)}$ & $0.16^{+0.01}_{-0.01}$ & $8.9^{+0.9}_{-0.8}$ &  & $0.38^{+0.01}_{-0.02}$ & $0.64^{+0.02}_{-0.02}$ & $3.7^{+0.2}_{-0.2}$ &  & $903 / 851$ \\
    E & $0$ to $15$ & $0.30 \textrm{ (fixed)}$ & $0.19^{+0.01}_{-0.01}$ & $2.2^{+0.3}_{-0.3}$ &  & $0.20^{+0.10}_{-0.11}$ & $0.81^{+0.04}_{-0.04}$ & $1.2^{+0.2}_{-0.2}$ &  & $774 / 761$ \\
    F & $15$ to $30$ & $0.088 \textrm{ (fixed)}$ & $0.19^{+0.01}_{-0.01}$ & $1.9^{+0.1}_{-0.1}$ &  & $\le 0.088$ & $0.79^{+0.04}_{-0.04}$ & $0.66^{+0.07}_{-0.07}$ &  & $931 / 883$ \\
    G & $30$ to $60$ & $0.024 \textrm{ (fixed)}$ & $0.19^{+0.01}_{-0.01}$ & $1.1^{+0.1}_{-0.1}$ &  & $\le 0.024$ & $0.72^{+0.02}_{-0.02}$ & $0.47^{+0.03}_{-0.03}$ &  & $961 / 883$ \\
    H & $60$ to $90$ & $0.014 \textrm{ (fixed)}$ & $0.19^{+0.01}_{-0.01}$ & $1.0^{+0.1}_{-0.1}$ &  & $\le 0.014$ & $0.66^{+0.04}_{-0.04}$ & $0.44^{+0.04}_{-0.04}$ &  & $952 / 883$ \\
    \enddata
    \tablecomments{All stacked regions are restricted to the Galactic longitude range of $60^\circ \leq l \leq 180^\circ$. Individual fields containing MAXI sources or the Cygnus superbubble were excluded from the stacked spectra. Quoted central values and statistical uncertainties are the posterior medians and 16th--84th percentiles. Upper limits are the 84th percentiles.}
    \tablenotetext{a}{Absorption column density in units of $10^{22}$ cm$^{-2}$.}
    \tablenotetext{b}{Plasma temperature in units of keV.}
    \tablenotetext{c}{Emission measure of the warm-hot component in units of $10^{-2}$ cm$^{-6}$ pc.}
    \tablenotetext{d}{Emission measure of the hot component in units of $10^{-3}$ cm$^{-6}$ pc.}
\end{deluxetable*}

\section{Discussion} \label{sec:discussion}

The individual-field and stacked analyses support a widespread two-temperature description of the Galactic soft X-ray emission.
Motivated by the structure-dependent EM correlation discussed in Section~\ref{sec:results}, we applied a unified spatial-modeling framework to the EM distributions of both the warm-hot and hot components to assess the relative contributions of stellar emission and diffuse plasma.

For the stellar contribution, we used the dM-star model of \citet{Wulf+2019}, while for the diffuse plasma contribution we adopted the empirical disk-like model and the adiabatic halo model of \citet{Kaaret+2020}.
In the dM-star model, the spatial distributions of young and old stars are represented as
\begin{align}
  \rho_{\mathrm{young}} &= \rho_0 \left(
    e^{-a^2/K_+^2} - e^{-a^2/K_-^2}
  \right), \\
  \rho_{\mathrm{old}} &= \rho_0 \left(
    e^{-\sqrt{0.5^2 + a^2/K_+^2}}
    - e^{-\sqrt{0.5^2 + a^2/K_-^2}}
  \right),
\end{align}
where $R$ and $z$ are the radial and vertical coordinates in the Galactocentric cylindrical coordinate system, and $a^2=R^2+z^2/c^2$ with the axis ratio $c$.
Here, $\rho$ represents the EM density associated with the stellar component, and the stellar EM along each line of sight was calculated as $\int \rho ds$.
We adopted the values of $\rho_0$, $K_+$, $K_-$, and $c$ for each stellar population from Table 10 of \citet{Wulf+2019}.
The total dM-star EM-density distribution was defined as the sum of the two stellar populations.
Because the dM-star model assumes $Z=1.0\,Z_{\odot}$, an abundance scaling factor of 0.3 is required for comparison with the warm-hot component, for which $Z=0.3\,Z_{\odot}$ was adopted.
The resulting dM-star model was used as a fixed spatial EM template.
No additional free normalization parameter was introduced in the fit.

For the disk-like component, we adopted
\begin{equation}
n(R,z)=
\begin{cases}
n_0 e^{-|z|/z_0}, & R \le R_c, \\[4pt]
n_0 e^{-(R-R_c)/R_0} e^{-|z|/z_0}, & R > R_c.
\end{cases}
\end{equation}
For the halo-like component, following \citet{Fang+2013}, we adopted the density profile
\begin{equation}
n(r)=n_{\rm v}
\left[
1+\frac{3.7}{x}\ln(1+x)
-\frac{3.7}{C_{\rm v}}\ln(1+C_{\rm v})
\right]^{3/2},
\end{equation}
where $x=r/R_{\rm s}$ and $C_{\rm v}=R_{\rm v}/R_{\rm s}$.
Here, $R$ and $z$ are as defined above, $r$ is the spherical radius from the Galactic center, and $n$ is the plasma density.
$R_{\rm s}$, $R_{\rm v}$, and $C_{\rm v}$ are the scale radius, virial radius, and concentration, respectively.
The EM of the diffuse plasma component along each line of sight was calculated as $\int n^2 ds$.
Following \citet{Kaaret+2020}, we fixed $R_c=5.5$~kpc and $R_0=2.0$~kpc for the disk-like component. For the halo-like component, we adopted $C_{\rm v}=12$ and $R_{\rm v}=260$~kpc following \citet{Fang+2013} and \citet{Kaaret+2020}.
The parameters describing the hybrid model are the scale height $z_0$ of the disk-like component, the central density $n_0$, and the virial-radius density $n_{\rm v}$ of the halo-like component.

Because the observed EMs can deviate from a smooth spatial distribution owing to unresolved sources, local foreground structures, and spatial variations averaged within the wide \textit{HaloSat} field of view, we introduced a patchiness parameter, $\sigma_{\rm p}$, in log-EM space, following the approach of \citet{Kaaret+2020} and similar treatments of intrinsic dispersion in CGM modeling \citep[e.g.,][]{Yang+2025}.
This parameter represents the average residual scatter of the observed EMs around the adopted smooth model, rather than an additional measurement uncertainty or a distinct physical component.
We fitted the spatial model with MCMC using the Goodman--Weare algorithm implemented in \texttt{emcee} \citep{Foreman-Mackey+2013}, adopting the same numbers of walkers, steps, and burn-in steps as in the spectral fitting.
The fit was evaluated with a $\chi^2$ statistic in log-EM space, including both the observational uncertainties and $\sigma_{\rm p}$.
To reduce the effects of bright X-ray sources and large-scale Galactic structures, we fitted 178 fields after excluding fields containing MAXI sources, the Cygnus and Orion--Eridanus superbubbles, and the eROSITA bubbles.

\begin{figure}[htbp]
\begin{center}
    \includegraphics[width=1.0\linewidth]{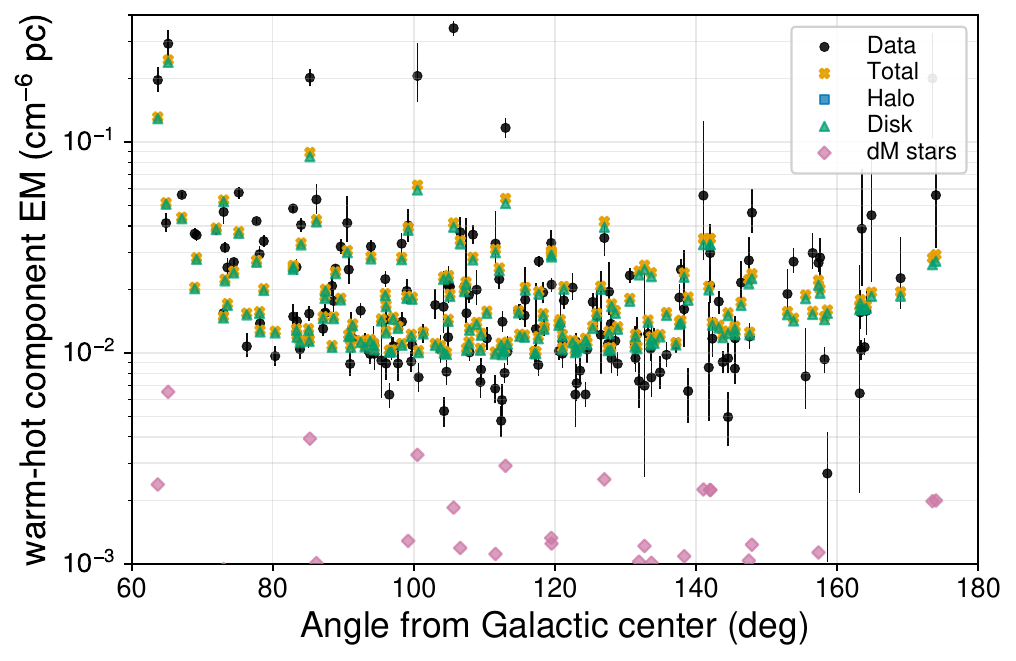}
    \includegraphics[width=1.0\linewidth]{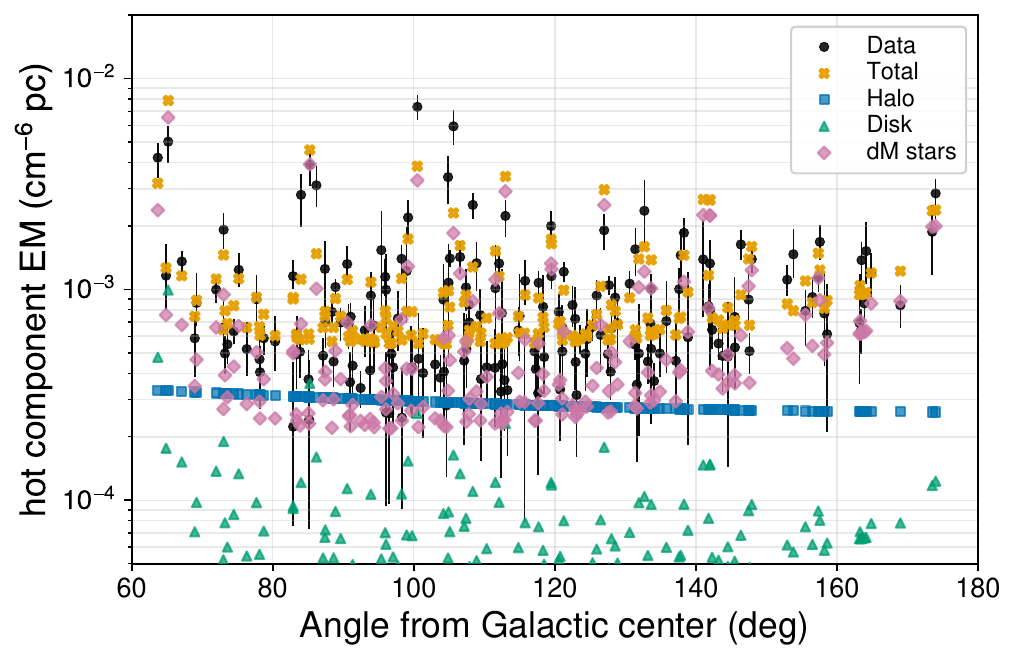}
\end{center}
    \caption{Spatial fits to the EM distributions after excluding regions containing bright X-ray sources and large-scale Galactic structures. Black circles with error bars show the observed EMs. Orange X-shaped markers show the total model, while blue squares, green triangles, and purple diamonds show the halo-like, disk-like, and dM-star contributions, respectively. The upper and lower panels correspond to the warm-hot and hot components, respectively. The halo-like component in the upper panel is not visible because it is constrained only as an upper limit.}
    \label{fig:em_fit}
\end{figure}

The resulting parameters are summarized in Table~\ref{tab:spatial-fit}, and the corresponding model profiles are shown in Figure~\ref{fig:em_fit}.
As a consistency check, we repeated the spatial fitting using alternative robust treatments, including a Huber-loss fit and a Student-$t$ likelihood, without the patchiness parameter.
We also repeated the fitting with alternative region selections by varying the exclusion of the Orion--Eridanus superbubble and the eROSITA bubbles.
The fitted parameters remained consistent within their uncertainties, and the main conclusions regarding the relative contributions of the stellar, disk-like, and halo-like components were unchanged.
\begin{deluxetable}{lcc}
\tablecaption{Spatial fitting results for the hybrid model.}
\label{tab:spatial-fit}
\tabletypesize{\footnotesize}
\tablewidth{0pt}
\tablehead{
\colhead{Parameter} &
\colhead{warm-hot component} &
\colhead{hot component}
}
\startdata
$n_{\rm v}~({\rm cm}^{-3})$ &
$<2.8\times10^{-6}$ &
$(1.4^{+0.2}_{-0.4})\times10^{-5}$ \\
$n_0~({\rm cm}^{-3})$ &
$(1.8^{+0.1}_{-0.1})\times10^{-2}$ &
$<2.2\times10^{-3}$ \\
$z_0~({\rm kpc})$ &
$0.75^{+0.12}_{-0.11}$ &
$<1.2$ \\
$\sigma_{\rm p}~({\rm dex})$ &
$0.17^{+0.01}_{-0.01}$ &
$0.12^{+0.01}_{-0.01}$ \\
\enddata
\tablecomments{The fit includes 178 fields after excluding regions containing MAXI sources, the Cygnus and Orion--Eridanus superbubbles, and the eROSITA bubbles.
The parameter $n_{\rm v}$ is the density of the halo-like component at the virial radius, $n_0$ is the central density of the disk-like component, $z_0$ is the disk scale height, and $\sigma_{\rm p}$ is the patchiness parameter.
Upper limits correspond to the 84th percentiles of the posterior distributions.
}
\end{deluxetable}

For the warm-hot component, $n_0$ and $z_0$ were well constrained, whereas $n_{\rm v}$ was constrained only as an upper limit.
Within the adopted hybrid model, the observed EM distribution is therefore described mainly by disk-like diffuse plasma, with a minor stellar contribution.
The present framework includes a fixed stellar EM template, which was not included in the model of \citet{Kaaret+2020}.
The upper limit on $n_{\rm v}$ is more than an order of magnitude below the virial-radius density reported by \citet{Kaaret+2020}, while $n_0$ is about twice as high and $z_0$ is about half as large as the corresponding values in that study.
Over the fields included in the spatial fitting, the halo-like term produces a comparatively uniform EM contribution, whereas the observed spatial variation is dominated by the disk-like morphology.
This result does not rule out tenuous extended hot gas because X-ray emission is weighted by the density squared and is preferentially sensitive to denser plasma, whereas absorption measurements depend approximately linearly on density and can remain sensitive to low-density gas over long path lengths \citep[e.g.,][]{Nakashima+2018, Kaaret+2020}.
To compare the density normalizations on a common abundance scale, we converted the present value of $n_0$ to $Z=1.0\,Z_{\odot}$ using $n_0\propto Z^{-1/2}$, obtaining $(9.9^{+0.5}_{-0.5})\times10^{-3}~\mathrm{cm^{-3}}$.
This value is about three times higher than that reported by \citet{Nakashima+2018} and about twice that reported by \citet{Ueda+2022}.
The present scale height is about half that reported by \citet{Nakashima+2018} and about three times larger than the value fixed by \citet{Ueda+2022}.
These differences may reflect differences in the adopted models, sky coverage, and field selection.

For the hot component, $n_0$ and $z_0$ were not independently constrained because of their degeneracy, whereas $n_{\rm v}$ was relatively well constrained.
The fit therefore did not require a significant disk-like contribution, although it did not exclude one.
To examine the allowed disk-like contribution, we repeated the fit with $z_0$ fixed at 0.75~kpc, the best-fitting value for the warm-hot component.
This fit yielded $n_0=(1.9^{+0.5}_{-0.8})\times10^{-3}~\mathrm{cm^{-3}}$, indicating that a nonzero disk-like normalization is allowed.
The resulting $n_0$ and the adopted $z_0$ are consistent within the uncertainties with the effective density normalization and scale height reported by \citet{Sugiyama+2023}, although the two studies adopted different spatial profiles and density definitions.
Within the adopted hybrid model, the EM distribution is described mainly by the dM-star model and the halo-like component, although a disk-like contribution remains possible.
This result is consistent with a hybrid picture in which stellar emission provides a major contribution to the hot emission and diffuse plasma may also contribute \citep[e.g.,][]{Ponti+2026}.
However, the halo-like component obtained here does not necessarily represent an extended, volume-filling super-virial CGM analogous to that commonly discussed for the warm-hot plasma.
As shown in Figure~\ref{fig:all-sky_results}, the hot component exhibits appreciable temperature variations, and the halo-like term may partly account for local hot plasma along different lines of sight or systematic uncertainties in the stellar component.
When evaluated at 10~kpc, the halo-like model yields a density of approximately $8\times10^{-5}~\mathrm{cm^{-3}}$, which remains below the upper limit on the super-virial halo density reported by \citet{Ponti+2026}, although the physical interpretation of the halo-like term remains model dependent.

The patchiness parameters were $\sigma_{\rm p}=0.17$ dex for the warm-hot component and $\sigma_{\rm p}=0.12$ dex for the hot component.
These values correspond to average residual scatters of factors of 1.5 and 1.3 around the adopted smooth spatial model, respectively.
The observed EM distributions therefore show a modest level of residual scatter, even after the stellar, disk-like, and halo-like components are included.

Using the fitted density distributions, we estimated the diffuse-plasma masses by integrating the disk-like component over all space and the halo-like component out to 260~kpc \citep{Kaaret+2020}.
For the warm-hot component, the disk-like mass was $M_{\rm disk}=7.8^{+0.8}_{-0.7}\times10^{7}~M_{\odot}$, and the halo-like mass was $M_{\rm halo}<4.2\times10^{9}~M_{\odot}$.
The disk-like mass is about four times that reported by \citet{Kaaret+2020}, about half the estimate of \citet{Nakashima+2018} after scaling to $Z=0.3Z_{\odot}$, and about 20\% lower than that of \citet{Ueda+2022}.
Compared with \citet{Kaaret+2020}, the present central density is about twice as high, whereas the scale height is about half as large.
These effects partly offset each other in the integrated mass, and the remaining difference likely reflects differences in model geometry, integration volume, and the treatment of stellar emission.
The halo-like mass was constrained only as an upper limit.
Because this estimate is derived from an emission-constrained density profile, it should not be interpreted as a general constraint on the total mass of low-density hot gas in the Galactic halo.
This limit remains compatible with an extended component at the $\sim10^{9}~M_{\odot}$ level discussed by \citet{Nakashima+2018} and is well below the halo mass of $(5.5$--$8.6)\times10^{10}~M_{\odot}$ inferred by \citet{Kaaret+2020}.
It is also consistent with the upper limit of a few $\times10^{10}~M_{\odot}$ reported by \citet{Ueda+2022} for a spherical component integrated out to 250~kpc at $Z=0.3Z_{\odot}$.
For the hot component, we obtained $M_{\rm disk}<9.6\times10^{6}~M_{\odot}$ and $M_{\rm halo}=2.0^{+0.2}_{-0.6}\times10^{10}~M_{\odot}$.
To our knowledge, these provide the first diffuse-plasma mass estimates for the hot component based on a hybrid spatial fit.
These estimates quantify the diffuse contribution after accounting for the substantial stellar contribution to the hot emission, although their absolute values remain dependent on the adopted stellar template.

To examine whether the disk-like diffuse plasma inferred from the spatial fitting is energetically plausible, we estimated its internal energy from the fitted density distributions.
The resulting energies are on the order of $\sim10^{56}$~erg for the warm-hot component and $\lesssim10^{55}$~erg for the hot component, comparable to or below the $\sim10^{56}$~erg energy associated with past Galactic-center activity \citep{Kataoka+2018} and that estimated for the eROSITA bubbles \citep{Predehl+2020}.
Thus, from the viewpoint of the total energy budget, disk-related feedback processes could plausibly contribute to these components, although detailed heating, cooling, and dissipation processes are required to identify the formation and maintenance mechanisms.
Although the disk-like parameters were independently constrained only for the warm-hot component, the spatial fits do not exclude a disk-related contribution to the hot component.
The two observed components may therefore represent a phenomenological decomposition of disk-related plasma with a continuous temperature distribution, rather than two completely independent physical phases \citep[e.g.,][]{Dutta+2024}.
This interpretation is also consistent with simulations and observations of external disk galaxies that suggest broad temperature distributions in diffuse X-ray-emitting gas \citep[e.g.,][]{Vijayan+2021, Wang+2021}.
Testing this possibility requires spectral models that explicitly include continuous temperature distributions and observations with higher energy resolution.

\section{Summary and Conclusions}\label{sec:summary-conclusions}

We analyzed the all-sky \textit{HaloSat} archive together with \textit{ROSAT} R1/R2 constraints to characterize soft X-ray emission associated with the Milky Way, extending our previous study of five Galactic-plane regions to 330 fields across the sky.
Uniform spectral fitting included the LHB, the warm-hot and hot components, SWCX, and the CXB.

Individual-field fits favored the two-temperature model at $>3\sigma$ in 60\% of the fields.
The warm-hot and hot components had median temperatures of $kT=0.19$ and $0.74$~keV, with 16th--84th percentile ranges of $0.17$--$0.21$ and $0.61$--$0.90$~keV, respectively.
The corresponding EM ranges were $(1.0$--$8.2)\times10^{-2}$ and $(0.5$--$4.1)\times10^{-3}~\mathrm{cm^{-6}\,pc}$.
As an ancillary result, the individual-field spectral fits yielded an all-sky LHB EM map based on spectral-component decomposition.
All eight stacked Galactic-latitude regions within $60^\circ\leq l\leq180^\circ$ favored the two-temperature model at $>3\sigma$ and reproduced the temperature and EM trends found in the individual fields.
The EMs of both components were approximately eight times higher at low than at high Galactic latitudes.
A single-temperature model with variable Ne remained disfavored relative to the standard two-temperature model.
A two-temperature model with variable N and Ne abundances improved on the standard model at $>3\sigma$ in only three stacked regions, with both abundances enhanced in only one region.
These results support a widespread two-temperature description of the Galactic soft X-ray sky.

The EM correlation between the two components weakened after fields associated with high-EM Galactic structures were excluded, indicating that the all-sky correlation does not necessarily imply a common physical origin.
We therefore excluded fields containing bright X-ray sources or large-scale Galactic structures and applied a hybrid spatial model comprising stellar emission and disk-like and halo-like diffuse plasma to the remaining 178 fields.
The warm-hot EM distribution was described mainly by the disk-like term, while the halo-like normalization was constrained only as an upper limit.
The hot distribution was described mainly by stellar emission with an additional halo-like term.
Because $n_0$ and $z_0$ were degenerate, the free fit did not require a disk-like contribution to the hot component, although such a contribution remained allowed.
Although the inferred relative contributions remain model dependent, the spatial decomposition suggests that the hot emission may have a composite stellar and diffuse origin.

The inferred diffuse-plasma masses were approximately $M_{\rm disk}=8\times10^{7}~M_{\odot}$ and $M_{\rm halo}<4\times10^{9}~M_{\odot}$ for the warm-hot component and $M_{\rm disk}<1\times10^{7}~M_{\odot}$ and $M_{\rm halo}=2\times10^{10}~M_{\odot}$ for the hot component.
These values should be regarded as model-dependent reference estimates because they depend on the adopted spatial geometry, integration volume, abundance, and stellar template.
The corresponding disk-like internal energies were of order $10^{56}$ and $\lesssim10^{55}$~erg, respectively, and are compatible with the energy available from Galactic feedback.
In this context, the two fitted temperatures may provide a phenomenological approximation to a broader underlying temperature distribution rather than representing completely independent plasma phases.

\begin{acknowledgments}

This study was financially supported by Grants-in-Aid for Scientific Research (KAKENHI) of the Japanese Society for the Promotion of Science (JSPS) under grant Nos. JP25K01012 and JP22K18274 and JST FOREST Program (Grant Number JPMJFR2369, Japan) awarded to I.M., and by NASA grant 80NSSC22K0624, ``Final Archive of the HaloSat Data.''
K.A. acknowledges the generous support of the Hattori International Scholarship Foundation (HISF) and the ``THERS Make New Standards Program for the Next Generation Researchers.''
D.K. acknowledges financial support from the Centre national d'\'etudes spatiales (CNES), France (ROR: \url{https://ror.org/04h1h0y33}) within the framework of the HaloSat mission.
ChatGPT (OpenAI) was used for English translation and language editing, with all generated text reviewed and revised by the authors.

\end{acknowledgments}

\bibliographystyle{aasjournal}
\bibliography{ref_cgm_halosat}

@ARTICLE{Tumlinson+2017,
       title           = {The Circumgalactic Medium},
       author          = {Tumlinson, Jason and Peeples, Molly S. and Werk, Jessica K.},
       journal         = {\araa},
       volume          = {55},
       number          = {1},
       pages           = {389--432},
       doi             = {10.1146/annurev-astro-091916-055240},
       url             = {https://doi.org/10.1146/annurev-astro-091916-055240},
       publisher       = {Annual Reviews},
       year            = {2017},
       month           = {August},
}

@ARTICLE{Putman+2012,
       title           = {Gaseous Galaxy Halos},
       author          = {Putman, M.E. and Peek, J.E.G. and Joung, M.R.},
       journal         = {\araa},
       volume          = {50},
       number          = {1},
       pages           = {491--529},
       doi             = {10.1146/annurev-astro-081811-125612},
       url             = {https://doi.org/10.1146/annurev-astro-081811-125612},
       publisher       = {Annual Reviews},
       year            = {2012},
       month           = {September},
}

@INCOLLECTION{Chen+2026,
       title           = {The circumgalactic medium},
       author          = {Chen, Hsiao-Wen and Zahedy, Fakhri S.},
       booktitle         = {Encyclopedia of Astrophysics},
       pages           = {370--400},
       doi             = {10.1016/b978-0-443-21439-4.00059-6},
       url             = {https://doi.org/10.1016/b978-0-443-21439-4.00059-6},
       publisher       = {Elsevier},
       year            = {2026},
}

@ARTICLE{Yoshino+2009,
       title           = {Energy Spectra of the Soft X-Ray Diffuse Emission in Fourteen Fields Observed with Suzaku},
       author          = {Yoshino, Tomotaka and Mitsuda, Kazuhiasa and Yamasaki, Noriko Y. and Takei, Yoh and Hagihara, Toshishige and Masui, Kensuke and Bauer, Michael and McCammon, Dan and Fujimoto, Ryuichi and Daniel Wang, Q. and Yao, Yangsen},
       journal         = {\pasj},
       volume          = {61},
       number          = {4},
       pages           = {805--823},
       doi             = {10.1093/pasj/61.4.805},
       url             = {https://doi.org/10.1093/pasj/61.4.805},
       publisher       = {Oxford University Press (OUP)},
       year            = {2009},
       month           = {August},
}

@ARTICLE{Gatuzz+2017,
       title           = {Probing the structure of the gas in the Milky Way through X-ray high-resolution spectroscopy},
       author          = {Gatuzz, Efra{\'i}n and Churazov, Eugene},
       journal         = {\mnras},
       volume          = {474},
       number          = {1},
       pages           = {696--711},
       doi             = {10.1093/mnras/stx2776},
       url             = {https://doi.org/10.1093/mnras/stx2776},
       publisher       = {Oxford University Press (OUP)},
       year            = {2018},
       month           = {February},
}

@ARTICLE{Kaaret+2020,
       title           = {A disk-dominated and clumpy circumgalactic medium of the Milky Way seen in X-ray emission},
       author          = {Kaaret, P. and Koutroumpa, D. and Kuntz, K. D. and Jahoda, K. and Bluem, J. and Gulick, H. and Hodges-Kluck, E. and LaRocca, D. M. and Ringuette, R. and Zajczyk, A.},
       journal         = {Nature Astronomy},
       volume          = {4},
       number          = {11},
       pages           = {1072--1077},
       doi             = {10.1038/s41550-020-01215-w},
       url             = {https://doi.org/10.1038/s41550-020-01215-w},
       publisher       = {Springer Science and Business Media LLC},
       year            = {2020},
       month           = {October},
}

@ARTICLE{Mathur+2021,
       month           = {February},
       year            = {2021},
       publisher       = {American Astronomical Society},
       url             = {https://doi.org/10.3847/1538-4357/abd03f},
       doi             = {10.3847/1538-4357/abd03f},
       pages           = {69},
       number          = {1},
       volume          = {908},
       journal         = {\apj},
       author          = {Mathur, Smita and Gupta, Anjali and Das, Sanskriti and Krongold, Yair and Nicastro, Fabrizio},
       title           = {Probing the Hot Circumgalactic Medium with Broad O VI and X-Rays},
}

@ARTICLE{Gupta+2021,
       title           = {Supervirial Temperature or Neon Overabundance? Suzaku Observations of the Milky Way Circumgalactic Medium},
       author          = {Gupta, Anjali and Kingsbury, Joshua and Mathur, Smita and Das, Sanskriti and Galeazzi, Massimiliano and Krongold, Yair and Nicastro, Fabrizio},
       journal         = {\apj},
       volume          = {909},
       number          = {2},
       pages           = {164},
       doi             = {10.3847/1538-4357/abdbb6},
       url             = {https://doi.org/10.3847/1538-4357/abdbb6},
       publisher       = {American Astronomical Society},
       year            = {2021},
       month           = {March},
}

@ARTICLE{Gupta+2023,
       month           = {May},
       year            = {2023},
       publisher       = {Springer Science and Business Media LLC},
       url             = {https://doi.org/10.1038/s41550-023-01963-5},
       doi             = {10.1038/s41550-023-01963-5},
       pages           = {799},
       number          = {7},
       volume          = {7},
       journal         = {Nature Astronomy},
       author          = {Gupta, Anjali and Mathur, Smita and Kingsbury, Joshua and Das, Sanskriti and Krongold, Yair},
       title           = {Thermal and chemical properties of the eROSITA bubbles from Suzaku observations},
}

@ARTICLE{Gupta+2025,
       title           = {Where Is the Supervirial Gas? III. Insights from X-Ray Shadow Observations and a Revised Model for the Soft Diffuse X-Ray Background},
       author          = {Gupta, Anjali and Mathur, Smita and Kingsbury, Joshua and Korkmaz, Esma and Das, Sanskriti and Krongold, Yair and Roy, Manami and Lara-DI, Armando},
       journal         = {\apj},
       volume          = {989},
       number          = {2},
       pages           = {194},
       doi             = {10.3847/1538-4357/adf10b},
       url             = {https://doi.org/10.3847/1538-4357/adf10b},
       publisher       = {American Astronomical Society},
       year            = {2025},
       month           = {August},
}

@ARTICLE{Das+2019a,
       title           = {Discovery of a Very Hot Phase of the Milky Way Circumgalactic Medium with Non-solar Abundance Ratios},
       author          = {Das, Sanskriti and Mathur, Smita and Nicastro, Fabrizio and Krongold, Yair},
       journal         = {\apjl},
       volume          = {882},
       number          = {2},
       pages           = {L23},
       doi             = {10.3847/2041-8213/ab3b09},
       url             = {https://doi.org/10.3847/2041-8213/ab3b09},
       publisher       = {American Astronomical Society},
       year            = {2019},
       month           = {September},
}

@ARTICLE{Bhattacharyya+2023,
       title           = {The Hot Circumgalactic Medium of the Milky Way: New Insights from XMM-Newton Observations},
       author          = {Bhattacharyya, Joy and Das, Sanskriti and Gupta, Anjali and Mathur, Smita and Krongold, Yair},
       journal         = {\apj},
       volume          = {952},
       number          = {1},
       pages           = {41},
       doi             = {10.3847/1538-4357/acd337},
       url             = {https://doi.org/10.3847/1538-4357/acd337},
       publisher       = {American Astronomical Society},
       year            = {2023},
       month           = {July},
}

@ARTICLE{Sugiyama+2023,
       title           = {The soft X-ray background with Suzaku. II. Supervirial temperature bubbles?},
       author          = {Sugiyama, Hayato and Ueda, Masaki and Fukushima, Kotaro and Kobayashi, Shogo B and Yamasaki, Noriko Y and Sato, Kosuke and Matsushita, Kyoko},
       journal         = {\pasj},
       volume          = {75},
       number          = {6},
       pages           = {1324--1336},
       doi             = {10.1093/pasj/psad073},
       url             = {https://doi.org/10.1093/pasj/psad073},
       publisher       = {Oxford University Press (OUP)},
       year            = {2023},
       month           = {December},
}

@ARTICLE{Masui+2009,
       title           = {The Nature of Unresolved Soft X-Ray Emission from the Galactic Disk},
       author          = {Masui, Kensuke and Mitsuda, Kazuhisa and Yamasaki, Noriko Y. and Takei, Yoh and Kimura, Shunsuke and Yoshino, Tomotaka and McCammon, Dan},
       journal         = {\pasj},
       volume          = {61},
       number          = {sp1},
       pages           = {S115--S122},
       doi             = {10.1093/pasj/61.sp1.s115},
       url             = {https://doi.org/10.1093/pasj/61.sp1.s115},
       publisher       = {Oxford University Press (OUP)},
       year            = {2009},
       month           = {January},
}

@ARTICLE{Wulf+2019,
       title           = {A High Spectral Resolution Study of the Soft X-Ray Background with the X-Ray Quantum Calorimeter},
       author          = {Wulf, Dallas and Eckart, Megan E and Galeazzi, Massimiliano and Jaeckel, Felix and Kelley, Richard L and Kilbourne, Caroline A and Morgan, Kelsey M and Porter, F Scott and McCammon, Dan and Szymkowiak, Andrew E},
       journal         = {\apj},
       volume          = {884},
       number          = {2},
       pages           = {120},
       doi             = {10.3847/1538-4357/ab41f8},
       url             = {https://doi.org/10.3847/1538-4357/ab41f8},
       publisher       = {American Astronomical Society},
       year            = {2019},
       month           = {October},
}

@ARTICLE{Ampuku+2024,
       title           = {Soft X-Ray Energy Spectra in the Wide-field Galactic Disk Area Revealed with HaloSat},
       author          = {Ampuku, Kazuki and Mitsuishi, Ikuyuki and Sakuta, Koki and Kaaret, Philip and LaRocca, Daniel M. and Angelini, Lorella},
       journal         = {\apj},
       volume          = {962},
       number          = {2},
       pages           = {153},
       doi             = {10.3847/1538-4357/ad1240},
       url             = {https://doi.org/10.3847/1538-4357/ad1240},
       publisher       = {American Astronomical Society},
       year            = {2024},
       month           = {February},
}

@ARTICLE{Knies+2024,
       title           = {A new understanding of the Gemini-Monoceros X-ray enhancement from discoveries with eROSITA},
       author          = {Knies, J. R. and Sasaki, M. and Becker, W. and Liu, T. and Ponti, G. and Plucinsky, P. P.},
       journal         = {\aap},
       volume          = {688},
       pages           = {A90},
       doi             = {10.1051/0004-6361/202348834},
       url             = {https://doi.org/10.1051/0004-6361/202348834},
       publisher       = {EDP Sciences},
       year            = {2024},
       month           = {August},
}

@ARTICLE{Zheng+2026,
       title           = {The average X-ray spectrum of the volume-complete M-, F-, G-, and K-type star sample within 10 pc of the Sun},
       author          = {Zheng, Xueying and Ponti, Gabriele and Locatelli, Nicola and Stelzer, Beate and Magaudda, Enza and Dennerl, Konrad and Freyberg, Michael and Sanders, Jeremy and Caramazza, Marilena and Sasaki, Manami and Merloni, Andrea and Robrade, Jan and Liu, Teng and Zhang, He-shou and Mayer, Martin G. F. and Zhang, Yi and Yeung, Michael C. H. and Becker, Werner},
       journal         = {\aap},
       doi             = {10.1051/0004-6361/202557865},
       url             = {https://doi.org/10.1051/0004-6361/202557865},
       publisher       = {EDP Sciences},
       year            = {2026},
       month           = {April},
       volume = {709},
       pages           = "A275",
}

@ARTICLE{Das+2021,
       title           = {The Hot Circumgalactic Medium of the Milky Way: Evidence for Supervirial, Virial, and Subvirial Temperatures; Nonsolar Chemical Composition; and Nonthermal Line Broadening},
       author          = {Das, Sanskriti and Mathur, Smita and Gupta, Anjali and Krongold, Yair},
       journal         = {\apj},
       volume          = {918},
       number          = {2},
       pages           = {83},
       doi             = {10.3847/1538-4357/ac0e8e},
       url             = {https://doi.org/10.3847/1538-4357/ac0e8e},
       publisher       = {American Astronomical Society},
       year            = {2021},
       month           = {September},
}

@ARTICLE{Roy+2025,
       title           = {Where is the Supervirial Gas? II. Insight from the Survey of Galactic Sightlines},
       author          = {Roy, Manami and Mathur, Smita and Das, Sanskriti and Lara-DI, Armando and Krongold, Yair and Gupta, Anjali},
       journal         = {\apj},
       volume          = {982},
       number          = {1},
       pages           = {8},
       doi             = {10.3847/1538-4357/adb3a5},
       url             = {https://doi.org/10.3847/1538-4357/adb3a5},
       publisher       = {American Astronomical Society},
       year            = {2025},
       month           = {March},
}

@ARTICLE{Bluem+2022,
       title           = {Widespread Detection of Two Components in the Hot Circumgalactic Medium of the Milky Way},
       author          = {Bluem, Jesse and Kaaret, Philip and Kuntz, K. D. and Jahoda, Keith M. and Koutroumpa, Dimitra and Hodges-Kluck, Edmund J. and Fuller, Chase A. and LaRocca, Daniel M. and Zajczyk, Anna},
       journal         = {\apj},
       volume          = {936},
       number          = {1},
       pages           = {72},
       doi             = {10.3847/1538-4357/ac8662},
       url             = {https://doi.org/10.3847/1538-4357/ac8662},
       publisher       = {American Astronomical Society},
       year            = {2022},
       month           = {September},
}

@ARTICLE{Ponti+2026,
       month           = {March},
       year            = {2026},
       publisher       = {EDP Sciences},
       url             = {https://doi.org/10.1051/0004-6361/202556925},
       doi             = {10.1051/0004-6361/202556925},
       pages           = {A320},
       volume          = {707},
       journal         = {\aap},
       author          = {Ponti, G. and Yeung, M. C. H. and Stel, G. and Locatelli, N. and Zheng, X. and Stelzer, B. and Merloni, A. and Caramazza, M. and Magaudda, E. and Sasaki, M. and Dennerl, K. and Reiprich, T. H. and Schwope, A. and Becker, W. and Freyberg, M.},
       title           = {Low-mass stars dominate the hot (0.7 keV) Galactic X-ray emission},
}

@ARTICLE{Kuntz+2019,
       title           = {Solar wind charge exchange: an astrophysical nuisance},
       author          = {Kuntz, K. D.},
       journal         = {\aapr},
       volume          = {27},
       number          = {1},
       pages           = {1--71},
       doi             = {10.1007/s00159-018-0114-0},
       url             = {https://doi.org/10.1007/s00159-018-0114-0},
       publisher       = {Springer Science and Business Media LLC},
       year            = {2019},
       month           = {December},
}

@ARTICLE{Kaaret+2019,
       title           = {HaloSat: A CubeSat to Study the Hot Galactic Halo},
       author          = {Kaaret, P. and Zajczyk, A. and LaRocca, D. M. and Ringuette, R. and Bluem, J. and Fuelberth, W. and Gulick, H. and Jahoda, K. and Johnson, T. E. and Kirchner, D. L. and Koutroumpa, D. and Kuntz, K. D. and McCurdy, R. and Miles, D. M. and Robison, W. T. and Silich, E. M.},
       journal         = {\apj},
       volume          = {884},
       number          = {2},
       pages           = {162},
       doi             = {10.3847/1538-4357/ab4193},
       url             = {https://doi.org/10.3847/1538-4357/ab4193},
       publisher       = {American Astronomical Society},
       year            = {2019},
       month           = {October},
}

@ARTICLE{Zajczyk+2020,
       title           = {On-ground calibration of the HaloSat science instrument},
       author          = {Zajczyk, Anna and Kaaret, Philip and LaRocca, Daniel and Fuelberth, William and Gulick, Hannah C. and Jahoda, Keith and Kirchner, Donald L. and McCurdy, Ross and Robison, William T. and Silich, Emily},
       journal         = {Journal of Astronomical Telescopes, Instruments, and Systems},
       volume          = {6},
       number          = {04},
       doi             = {10.1117/1.jatis.6.4.044005},
       url             = {https://doi.org/10.1117/1.jatis.6.4.044005},
       publisher       = {SPIE-Intl Soc Optical Eng},
       year            = {2020},
       month           = {December},
       pages           = {044005},
}

@ARTICLE{Silich+2021,
       title           = {A Search for the 3.5 keV Line from the Milky Way's Dark Matter Halo with HaloSat},
       author          = {Silich, E. M. and Jahoda, K. and Angelini, L. and Kaaret, P. and Zajczyk, A. and LaRocca, D. M. and Ringuette, R. and Richardson, J.},
       journal         = {\apj},
       volume          = {916},
       number          = {1},
       pages           = {2},
       doi             = {10.3847/1538-4357/ac043b},
       url             = {https://doi.org/10.3847/1538-4357/ac043b},
       publisher       = {American Astronomical Society},
       year            = {2021},
       month           = {July},
}

@ARTICLE{Ringuette+2021,
       title           = {HaloSat Observations of Heliospheric Solar Wind Charge Exchange},
       author          = {Ringuette, R. and Koutroumpa, D. and Kuntz, K. D. and Kaaret, P. and Jahoda, K. and LaRocca, D. and Kounkel, M. and Richardson, J. and Zajczyk, A. and Bluem, J.},
       journal         = {\apj},
       volume          = {918},
       number          = {2},
       pages           = {41},
       doi             = {10.3847/1538-4357/ac0e33},
       url             = {https://doi.org/10.3847/1538-4357/ac0e33},
       publisher       = {American Astronomical Society},
       year            = {2021},
       month           = {September},
}

@ARTICLE{Snowden+1997,
       month           = {August},
       year            = {1997},
       publisher       = {American Astronomical Society},
       url             = {https://doi.org/10.1086/304399},
       doi             = {10.1086/304399},
       pages           = {125--135},
       number          = {1},
       volume          = {485},
       journal         = {\apj},
       author          = {Snowden, S. L. and Egger, R. and Freyberg, M. J. and McCammon, D. and Plucinsky, P. P. and Sanders, W. T. and Schmitt, J. H. M. M. and Trumper, J. and Voges, W.},
       title           = {ROSAT Survey Diffuse X-Ray Background Maps. II.},
}

@ARTICLE{Yeung+2024,
       title           = {The SRG/eROSITA diffuse soft X-ray background},
       author          = {Yeung, Michael C. H. and Ponti, Gabriele and Freyberg, Michael J. and Dennerl, Konrad and Liu, Teng and Locatelli, Nicola and Mayer, Martin G. F. and Sanders, Jeremy S. and Sasaki, Manami and Strong, Andy and Zhang, Yi and Zheng, Xueying and Gatuzz, Efrain},
       journal         = {\aap},
       volume          = {690},
       pages           = {A399},
       doi             = {10.1051/0004-6361/202451045},
       url             = {https://doi.org/10.1051/0004-6361/202451045},
       publisher       = {EDP Sciences},
       year            = {2024},
       month           = {October},
}

@INPROCEEDINGS{Arnaud+1996,
       adsurl          = {https://ui.adsabs.harvard.edu/abs/1996ASPC..101...17A},
       pages           = {17},
       month           = {January},
       volume          = {101},
       series          = {Astronomical Society of the Pacific Conference Series},
       editor          = {{Jacoby}, George H. and {Barnes}, Jeannette},
       year            = {1996},
       booktitle       = {Astronomical Data Analysis Software and Systems V},
       title           = {{XSPEC: The First Ten Years}},
       author          = {{Arnaud}, K.~A.},
}

@MISC{Gordon+2021,
       adsurl          = {https://ui.adsabs.harvard.edu/abs/2021ascl.soft01014G},
       eprint          = {2101.014},
       archiveprefix   = {ascl},
       eid             = {ascl:2101.014},
       month           = {January},
       year            = {2021},
       howpublished    = {Astrophysics Source Code Library, record ascl:2101.014},
       title           = {{PyXspec: Python interface to XSPEC spectral-fitting program}},
       author          = {{Gordon}, Craig and {Arnaud}, Keith},
}

@ARTICLE{Smith+2001,
       title           = {Collisional Plasma Models with APEC/APED: Emission-Line Diagnostics of Hydrogen-like and Helium-like Ions},
       author          = {Smith, Randall K. and Brickhouse, Nancy S. and Liedahl, Duane A. and Raymond, John C.},
       journal         = {\apj},
       volume          = {556},
       number          = {2},
       pages           = {L91--L95},
       doi             = {10.1086/322992},
       url             = {https://doi.org/10.1086/322992},
       publisher       = {American Astronomical Society},
       year            = {2001},
       month           = {August},
}

@ARTICLE{Wilms+2000,
       title           = {On the Absorption of X-Rays in the Interstellar Medium},
       author          = {Wilms, J. and Allen, A. and McCray, R.},
       journal         = {\apj},
       volume          = {542},
       number          = {2},
       pages           = {914--924},
       doi             = {10.1086/317016},
       url             = {https://doi.org/10.1086/317016},
       publisher       = {American Astronomical Society},
       year            = {2000},
       month           = {October},
}

@ARTICLE{Verner+1996,
       title           = {Atomic Data for Astrophysics. II. New Analytic FITS for Photoionization Cross Sections of Atoms and Ions},
       author          = {Verner, D. A. and Ferland, G. J. and Korista, K. T. and Yakovlev, D. G.},
       journal         = {\apj},
       volume          = {465},
       pages           = {487},
       doi             = {10.1086/177435},
       url             = {https://doi.org/10.1086/177435},
       publisher       = {American Astronomical Society},
       year            = {1996},
       month           = {July},
}

@ARTICLE{Cash+1979,
       title           = {Parameter estimation in astronomy through application of the likelihood ratio},
       author          = {Cash, W.},
       journal         = {\apj},
       volume          = {228},
       pages           = {939},
       doi             = {10.1086/156922},
       url             = {https://doi.org/10.1086/156922},
       publisher       = {American Astronomical Society},
       year            = {1979},
       month           = {March},
}

@ARTICLE{Yeung+2023,
       title           = {SRG/eROSITA X-ray shadowing study of giant molecular clouds},
       author          = {Yeung, M. C. H. and Freyberg, M. J. and Ponti, G. and Dennerl, K. and Sasaki, M. and Strong, A.},
       journal         = {\aap},
       volume          = {676},
       pages           = {A3},
       doi             = {10.1051/0004-6361/202345867},
       url             = {https://doi.org/10.1051/0004-6361/202345867},
       publisher       = {EDP Sciences},
       year            = {2023},
       month           = {August},
}

@ARTICLE{Koutroumpa+2012,
       title           = {Update on modeling and data analysis of heliospheric solar wind charge exchange X-ray emission},
       author          = {Koutroumpa, D.},
       journal         = {Astronomische Nachrichten},
       volume          = {333},
       number          = {4},
       pages           = {341--346},
       doi             = {10.1002/asna.201211666},
       url             = {https://doi.org/10.1002/asna.201211666},
       publisher       = {Wiley},
       year            = {2012},
       month           = {April},
}

@ARTICLE{Gloeckler+1998,
       month           = {July},
       year            = {1998},
       publisher       = {Springer Science and Business Media LLC},
       url             = {https://doi.org/10.1023/a:1005036131689},
       doi             = {10.1023/a:1005036131689},
       pages           = {497--539},
       number          = {1-4},
       volume          = {86},
       journal         = {\ssr},
       author          = {Gloeckler, G. and Cain, J. and Ipavich, F.M. and Tums, E.O. and Bedini, P. and Fisk, L.A. and Zurbuchen, T.H. and Bochsler, P. and Fischer, J. and Wimmer-Schweingruber, R.F. and Geiss, J. and Kallenbach, R.},
       title           = {Investigation of the composition of solar and interstellar matter using solar wind and pickup ion measurements with SWICS and SWIMS on the ACE spacecraft},
}

@ARTICLE{Uprety+2016,
       title           = {SOLAR WIND CHARGE EXCHANGE CONTRIBUTION TO THE ROSAT ALL SKY SURVEY MAPS},
       author          = {Uprety, Y. and Chiao, M. and Collier, M. R. and Cravens, T. and Galeazzi, M. and Koutroumpa, D. and Kuntz, K. D. and Lallement, R. and Lepri, S. T. and Liu, W. and McCammon, D. and Morgan, K. and Porter, F. S. and Prasai, K. and Snowden, S. L. and Thomas, N. E. and Ursino, E. and Walsh, B. M.},
       journal         = {\apj},
       volume          = {829},
       number          = {2},
       pages           = {83},
       doi             = {10.3847/0004-637x/829/2/83},
       url             = {https://doi.org/10.3847/0004-637x/829/2/83},
       publisher       = {American Astronomical Society},
       year            = {2016},
       month           = {October},
}

@ARTICLE{Liu+2017,
       title           = {THE STRUCTURE OF THE LOCAL HOT BUBBLE},
       author          = {Liu, W. and Chiao, M. and Collier, M. R. and Cravens, T. and Galeazzi, M. and Koutroumpa, D. and Kuntz, K. D. and Lallement, R. and Lepri, S. T. and McCammon, D. and Morgan, K. and Porter, F. S. and Snowden, S. L. and Thomas, N. E. and Uprety, Y. and Ursino, E. and Walsh, B. M.},
       journal         = {\apj},
       volume          = {834},
       number          = {1},
       pages           = {33},
       doi             = {10.3847/1538-4357/834/1/33},
       url             = {https://doi.org/10.3847/1538-4357/834/1/33},
       publisher       = {American Astronomical Society},
       year            = {2017},
       month           = {January},
}

@ARTICLE{Kushino+2002,
       title           = {Study of the X-Ray Background Spectrum and Its Large-Scale Fluctuation with ASCA},
       author          = {Kushino, Akihiro and Ishisaki, Yoshitaka and Morita, Umeyo and Yamasaki, Noriko Y. and Ishida, Manabu and Ohashi, Takaya and Ueda, Yoshihiro},
       journal         = {\pasj},
       volume          = {54},
       number          = {3},
       pages           = {327--352},
       doi             = {10.1093/pasj/54.3.327},
       url             = {https://doi.org/10.1093/pasj/54.3.327},
       publisher       = {Oxford University Press (OUP)},
       year            = {2002},
       month           = {June},
}

@ARTICLE{Planck+2014,
       month           = {November},
       year            = {2014},
       publisher       = {EDP Sciences},
       url             = {https://doi.org/10.1051/0004-6361/201323195},
       doi             = {10.1051/0004-6361/201323195},
       pages           = {A11},
       volume          = {571},
       journal         = {\aap},
       author          = {{Planck Collaboration} and Abergel, A. and Ade, P. A. R. and Aghanim, N. and Alves, M. I. R. and Aniano, G. and Armitage-Caplan, C. and Arnaud, M. and Ashdown, M. and Atrio-Barandela, F. and Aumont, J. and Baccigalupi, C. and Banday, A. J. and Barreiro, R. B. and Bartlett, J. G. and Battaner, E. and Benabed, K. and Beno{\^i}t, A. and Benoit-L{\'e}vy, A. and Bernard, J.-P. and Bersanelli, M. and Bielewicz, P. and Bobin, J. and Bock, J. J. and Bonaldi, A. and Bond, J. R. and Borrill, J. and Bouchet, F. R. and Boulanger, F. and Bridges, M. and Bucher, M. and Burigana, C. and Butler, R. C. and Cardoso, J.-F. and Catalano, A. and Chamballu, A. and Chary, R.-R. and Chiang, H. C. and Chiang, L.-Y and Christensen, P. R. and Church, S. and Clemens, M. and Clements, D. L. and Colombi, S. and Colombo, L. P. L. and Combet, C. and Couchot, F. and Coulais, A. and Crill, B. P. and Curto, A. and Cuttaia, F. and Danese, L. and Davies, R. D. and Davis, R. J. and de Bernardis, P. and de Rosa, A. and de Zotti, G. and Delabrouille, J. and Delouis, J.-M. and D{\'e}sert, F.-X. and Dickinson, C. and Diego, J. M. and Dole, H. and Donzelli, S. and Dor{\'e}, O. and Douspis, M. and Draine, B. T. and Dupac, X. and Efstathiou, G. and En{\ss}lin, T. A. and Eriksen, H. K. and Falgarone, E. and Finelli, F. and Forni, O. and Frailis, M. and Fraisse, A. A. and Franceschi, E. and Galeotta, S. and Ganga, K. and Ghosh, T. and Giard, M. and Giardino, G. and Giraud-H{\'e}raud, Y. and Gonz{\'a}lez-Nuevo, J. and G{\'o}rski, K. M. and Gratton, S. and Gregorio, A. and Grenier, I. A. and Gruppuso, A. and Guillet, V. and Hansen, F. K. and Hanson, D. and Harrison, D. L. and Helou, G. and Henrot-Versill{\'e}, S. and Hern{\'a}ndez-Monteagudo, C. and Herranz, D. and Hildebrandt, S. R. and Hivon, E. and Hobson, M. and Holmes, W. A. and Hornstrup, A. and Hovest, W. and Huffenberger, K. M. and Jaffe, A. H. and Jaffe, T. R. and Jewell, J. and Joncas, G. and Jones, W. C. and Juvela, M. and Keih{\"a}nen, E. and Keskitalo, R. and Kisner, T. S. and Knoche, J. and Knox, L. and Kunz, M. and Kurki-Suonio, H. and Lagache, G. and L{\"a}hteenm{\"a}ki, A. and Lamarre, J.-M. and Lasenby, A. and Laureijs, R. J. and Lawrence, C. R. and Leonardi, R. and Le{\'o}n-Tavares, J. and Lesgourgues, J. and Levrier, F. and Liguori, M. and Lilje, P. B. and Linden-V{\o}rnle, M. and L{\'o}pez-Caniego, M. and Lubin, P. M. and Mac{\'i}as-P{\'e}rez, J. F. and Maffei, B. and Maino, D. and Mandolesi, N. and Maris, M. and Marshall, D. J. and Martin, P. G. and Mart{\'i}nez-Gonz{\'a}lez, E. and Masi, S. and Massardi, M. and Matarrese, S. and Matthai, F. and Mazzotta, P. and McGehee, P. and Melchiorri, A. and Mendes, L. and Mennella, A. and Migliaccio, M. and Mitra, S. and Miville-Desch{\^e}nes, M.-A. and Moneti, A. and Montier, L. and Morgante, G. and Mortlock, D. and Munshi, D. and Murphy, J. A. and Naselsky, P. and Nati, F. and Natoli, P. and Netterfield, C. B. and N{\o}rgaard-Nielsen, H. U. and Noviello, F. and Novikov, D. and Novikov, I. and Osborne, S. and Oxborrow, C. A. and Paci, F. and Pagano, L. and Pajot, F. and Paladini, R. and Paoletti, D. and Pasian, F. and Patanchon, G. and Perdereau, O. and Perotto, L. and Perrotta, F. and Piacentini, F. and Piat, M. and Pierpaoli, E. and Pietrobon, D. and Plaszczynski, S. and Pointecouteau, E. and Polenta, G. and Ponthieu, N. and Popa, L. and Poutanen, T. and Pratt, G. W. and Pr{\'e}zeau, G. and Prunet, S. and Puget, J.-L. and Rachen, J. P. and Reach, W. T. and Rebolo, R. and Reinecke, M. and Remazeilles, M. and Renault, C. and Ricciardi, S. and Riller, T. and Ristorcelli, I. and Rocha, G. and Rosset, C. and Roudier, G. and Rowan-Robinson, M. and Rubi{\~n}o-Mart{\'i}n, J. A. and Rusholme, B. and Sandri, M. and Santos, D. and Savini, G. and Scott, D. and Seiffert, M. D. and Shellard, E. P. S. and Spencer, L. D. and Starck, J.-L. and Stolyarov, V. and Stompor, R. and Sudiwala, R. and Sunyaev, R. and Sureau, F. and Sutton, D. and Suur-Uski, A.-S. and Sygnet, J.-F. and Tauber, J. A. and Tavagnacco, D. and Terenzi, L. and Toffolatti, L. and Tomasi, M. and Tristram, M. and Tucci, M. and Tuovinen, J. and T{\"u}rler, M. and Umana, G. and Valenziano, L. and Valiviita, J. and Van Tent, B. and Verstraete, L. and Vielva, P. and Villa, F. and Vittorio, N. and Wade, L. A. and Wandelt, B. D. and Welikala, N. and Ysard, N. and Yvon, D. and Zacchei, A. and Zonca, A.},
       title           = {{Planck} 2013 results. XI. All-sky model of thermal dust emission},
}

@ARTICLE{Zhu+2017,
       title           = {The gas-to-extinction ratio and the gas distribution in the Galaxy},
       author          = {Zhu, Hui and Tian, Wenwu and Li, Aigen and Zhang, Mengfei},
       journal         = {\mnras},
       volume          = {471},
       number          = {3},
       pages           = {3494--3528},
       doi             = {10.1093/mnras/stx1580},
       url             = {https://doi.org/10.1093/mnras/stx1580},
       publisher       = {Oxford University Press (OUP)},
       year            = {2017},
       month           = {November},
}

@ARTICLE{Tomida+2016,
       title           = {The first MAXI/SSC catalog of X-ray sources in 0.7–7.0 keV},
       author          = {Tomida, Hiroshi and Uchida, Daiki and Tsunemi, Hiroshi and Imatani, Ritsuko and Kimura, Masashi and Nakahira, Satoshi and Hanayama, Takanori and Yoshidome, Koshiro},
       journal         = {\pasj},
       volume          = {68},
       number          = {SP1},
       pages           = {S32},
       doi             = {10.1093/pasj/psw006},
       url             = {https://doi.org/10.1093/pasj/psw006},
       publisher       = {Oxford University Press (OUP)},
       year            = {2016},
       month           = {June},
}

@ARTICLE{Ueda+2022,
       title           = {The soft X-ray background with Suzaku. I. Milky Way halo},
       author          = {Ueda, Masaki and Sugiyama, Hayato and Kobayashi, Shogo B and Fukushima, Kotaro and Yamasaki, Noriko Y and Sato, Kosuke and Matsushita, Kyoko},
       journal         = {\pasj},
       volume          = {74},
       number          = {6},
       pages           = {1396--1414},
       doi             = {10.1093/pasj/psac077},
       url             = {https://doi.org/10.1093/pasj/psac077},
       publisher       = {Oxford University Press (OUP)},
       year            = {2022},
       month           = {December},
}

@ARTICLE{Rachford+2002,
       title           = {A Far Ultraviolet Spectroscopic Explorer Survey of Interstellar Molecular Hydrogen in Translucent Clouds},
       author          = {Rachford, Brian L. and Snow, Theodore P. and Tumlinson, Jason and Shull, J. Michael and Blair, William P. and Ferlet, Roger and Friedman, Scott D. and Gry, Cecile and Jenkins, Edward B. and Morton, Donald C. and Savage, Blair D. and Sonnentrucker, Paule and Vidal-Madjar, Alfred and Welty, Daniel E. and York, Donald G.},
       journal         = {\apj},
       volume          = {577},
       number          = {1},
       pages           = {221--244},
       doi             = {10.1086/342146},
       url             = {https://doi.org/10.1086/342146},
       publisher       = {American Astronomical Society},
       year            = {2002},
       month           = {September},
}

@ARTICLE{Willingale+2013,
       title           = {Calibration of X-ray absorption in our Galaxy},
       author          = {Willingale, R. and Starling, R. L. C. and Beardmore, A. P. and Tanvir, N. R. and O'Brien, P. T.},
       journal         = {\mnras},
       volume          = {431},
       number          = {1},
       pages           = {394--404},
       doi             = {10.1093/mnras/stt175},
       url             = {https://doi.org/10.1093/mnras/stt175},
       publisher       = {Oxford University Press (OUP)},
       year            = {2013},
       month           = {May},
}

@ARTICLE{HI4PI+2016,
       month           = {October},
       year            = {2016},
       publisher       = {EDP Sciences},
       url             = {https://doi.org/10.1051/0004-6361/201629178},
       doi             = {10.1051/0004-6361/201629178},
       pages           = {A116},
       volume          = {594},
       journal         = {\aap},
       author          = {{HI4PI Collaboration} and Ben Bekhti, N. and Fl{\"o}er, L. and Keller, R. and Kerp, J. and Lenz, D. and Winkel, B. and Bailin, J. and Calabretta, M. R. and Dedes, L. and Ford, H. A. and Gibson, B. K. and Haud, U. and Janowiecki, S. and Kalberla, P. M. W. and Lockman, F. J. and McClure-Griffiths, N. M. and Murphy, T. and Nakanishi, H. and Pisano, D. J. and Staveley-Smith, L.},
       title           = {{HI4PI}: A full-sky H{\sc i} survey based on {EBHIS} and {GASS}},
}

@ARTICLE{Fang+2013,
       title           = {ON THE HOT GAS CONTENT OF THE MILKY WAY HALO},
       author          = {Fang, Taotao and Bullock, James and Boylan-Kolchin, Michael},
       journal         = {\apj},
       volume          = {762},
       number          = {1},
       pages           = {20},
       doi             = {10.1088/0004-637x/762/1/20},
       url             = {https://doi.org/10.1088/0004-637x/762/1/20},
       publisher       = {American Astronomical Society},
       year            = {2013},
       month           = {January},
}

@ARTICLE{Yang+2025,
       title           = {Modelling the cool gas clumps in the circumgalactic medium},
       author          = {Yang, Hang and Qu, Zhijie and Bregman, Joel N and Ji, Li},
       journal         = {\mnras},
       volume          = {538},
       number          = {3},
       pages           = {1871--1883},
       doi             = {10.1093/mnras/staf407},
       url             = {https://doi.org/10.1093/mnras/staf407},
       publisher       = {Oxford University Press (OUP)},
       year            = {2025},
       month           = {March},
}

@ARTICLE{Foreman-Mackey+2013,
       month           = {March},
       year            = {2013},
       publisher       = {IOP Publishing},
       url             = {https://doi.org/10.1086/670067},
       doi             = {10.1086/670067},
       pages           = {306--312},
       number          = {925},
       volume          = {125},
       journal         = {\pasp},
       author          = {Foreman-Mackey, Daniel and Hogg, David W. and Lang, Dustin and Goodman, Jonathan},
       title           = {{emcee}: The {MCMC} Hammer},
}

@ARTICLE{Nakashima+2018,
       title           = {Spatial Distribution of the Milky Way Hot Gaseous Halo Constrained by Suzaku X-Ray Observations},
       author          = {Nakashima, Shinya and Inoue, Yoshiyuki and Yamasaki, Noriko and Sofue, Yoshiaki and Kataoka, Jun and Sakai, Kazuhiro},
       journal         = {\apj},
       volume          = {862},
       number          = {1},
       pages           = {34},
       doi             = {10.3847/1538-4357/aacceb},
       url             = {https://doi.org/10.3847/1538-4357/aacceb},
       publisher       = {American Astronomical Society},
       year            = {2018},
       month           = {July},
}

@ARTICLE{Kataoka+2018,
       month           = {February},
       year            = {2018},
       publisher       = {MDPI AG},
       url             = {https://doi.org/10.3390/galaxies6010027},
       doi             = {10.3390/galaxies6010027},
       pages           = {27},
       number          = {1},
       volume          = {6},
       journal         = {Galaxies},
       author          = {Kataoka, Jun and Sofue, Yoshiaki and Inoue, Yoshiyuki and Akita, Masahiro and Nakashima, Shinya and Totani, Tomonori},
       title           = {X-Ray and Gamma-Ray Observations of the Fermi Bubbles and NPS/Loop I Structures},
}

@ARTICLE{Predehl+2020,
       title           = {Detection of large-scale X-ray bubbles in the Milky Way halo},
       author          = {Predehl, P. and Sunyaev, R. A. and Becker, W. and Brunner, H. and Burenin, R. and Bykov, A. and Cherepashchuk, A. and Chugai, N. and Churazov, E. and Doroshenko, V. and Eismont, N. and Freyberg, M. and Gilfanov, M. and Haberl, F. and Khabibullin, I. and Krivonos, R. and Maitra, C. and Medvedev, P. and Merloni, A. and Nandra, K. and Nazarov, V. and Pavlinsky, M. and Ponti, G. and Sanders, J. S. and Sasaki, M. and Sazonov, S. and Strong, A. W. and Wilms, J.},
       journal         = {\nat},
       volume          = {588},
       number          = {7837},
       pages           = {227},
       doi             = {10.1038/s41586-020-2979-0},
       url             = {https://doi.org/10.1038/s41586-020-2979-0},
       publisher       = {Springer Science and Business Media LLC},
       year            = {2020},
       month           = {December},
}

@ARTICLE{Dutta+2024,
       month           = {June},
       year            = {2024},
       publisher       = {Oxford University Press (OUP)},
       url             = {https://doi.org/10.1093/mnras/stae977},
       doi             = {10.1093/mnras/stae977},
       pages           = {5117--5139},
       number          = {4},
       volume          = {531},
       journal         = {\mnras},
       author          = {Dutta, Alankar and Bisht, Mukesh Singh and Sharma, Prateek and Ghosh, Ritali and Roy, Manami and Nath, Biman B},
       title           = {Beyond radial profiles: using log-normal distributions to model the multiphase circumgalactic medium},
}

@ARTICLE{Vijayan+2021,
       title           = {X-ray spectra of circumgalactic medium around star-forming galaxies: connecting simulations to observations},
       author          = {Vijayan, Aditi and Li, Miao},
       journal         = {\mnras},
       volume          = {510},
       number          = {1},
       pages           = {568--580},
       doi             = {10.1093/mnras/stab3413},
       url             = {https://doi.org/10.1093/mnras/stab3413},
       publisher       = {Oxford University Press (OUP)},
       year            = {2021},
       month           = {December},
}

@ARTICLE{Wang+2021,
       month           = {November},
       year            = {2021},
       publisher       = {Oxford University Press (OUP)},
       url             = {https://doi.org/10.1093/mnras/stab2997},
       doi             = {10.1093/mnras/stab2997},
       pages           = {6155--6175},
       number          = {4},
       volume          = {508},
       journal         = {\mnras},
       author          = {Wang, Q Daniel and Zeng, Yuxuan and Bogd{\'a}n, {\'A}kos and Ji, Li},
       title           = {Deep Chandra observations of diffuse hot plasma in M83},
}

\appendix
\restartappendixnumbering

\section{Individual-field Spectral-fitting Results}
\label{app:individual_results}

Table~\ref{tab:best-fit} presents a sample of the spectral-fitting results for the individual fields.
The complete table containing results for all 330 fields is available in machine-readable format.

\begin{rotatetable*}
    \begin{deluxetable*}{lccccccccccccccccc}
    \tablecaption{
    Spectral-fitting results for the 330 individual fields.
    \label{tab:best-fit}
    }
    \tabletypesize{\footnotesize}
    \tablehead{ Obs.ID & & \multicolumn{2}{c}{Pointing Direction} & & Exposure & & \multicolumn{1}{c}{LHB} & & \multicolumn{3}{c}{warm-hot component} & & \multicolumn{3}{c}{hot component} & & $\chi^2$/DoF\\ \cline{3-4} \cline{8-8} \cline{10-12} \cline{14-16} & & \textit{l} (deg.) & \textit{b} (deg.) & & (ks) & & EM\tablenotemark{a} & & $N_{\textrm{H}}$\tablenotemark{b} & $kT$\tablenotemark{c} & EM\tablenotemark{d} & & $N_{\textrm{H}}$\tablenotemark{b} & $kT$\tablenotemark{c} & EM\tablenotemark{e} & & }
    \startdata
    HS0004 &  & $58.09$ & $87.96$ &  & $30.5$ &  & $5.5^{+1.0}_{-1.0}$ &  & $0.014 \textrm{ (fixed)}$ & $0.18^{+0.01}_{-0.01}$ & $1.2^{+0.2}_{-0.2}$ &  & $0.0076^{+0.0065}_{-0.0049}$ & $0.70 \textrm{ (fixed)}$ & $0.69^{+0.13}_{-0.14}$ &  & $545 / 526$ \\
    HS0006 &  & $150.58$ & $-13.26$ &  & $25.0$ &  & $1.8^{+0.7}_{-0.7}$ &  & $0.16 \textrm{ (fixed)}$ & $0.16^{+0.02}_{-0.02}$ & $1.8^{+0.7}_{-0.4}$ &  & $0.16^{+0.00}_{-0.01}$ & $0.70 \textrm{ (fixed)}$ & $3.6^{+0.2}_{-0.2}$ &  & $787 / 473$ \\
    HS0008 &  & $278.32$ & $-34.00$ &  & $9.9$ &  & $2.5^{+0.5}_{-0.5}$ &  & $0.090 \textrm{ (fixed)}$ & $0.19^{+0.01}_{-0.01}$ & $5.5^{+0.4}_{-0.4}$ &  & $0.083^{+0.007}_{-0.010}$ & $0.70 \textrm{ (fixed)}$ & $5.8^{+0.4}_{-0.5}$ &  & $507 / 291$ \\
    HS0009 &  & $254.27$ & $-6.22$ &  & $24.7$ &  & $3.4^{+1.1}_{-1.2}$ &  & $0.39 \textrm{ (fixed)}$ & $0.16^{+0.01}_{-0.01}$ & $35^{+3}_{-3}$ &  & $0.25^{+0.08}_{-0.10}$ & $0.78^{+0.03}_{-0.03}$ & $5.9^{+1.1}_{-1.1}$ &  & $411 / 397$ \\
    HS0011 &  & $189.06$ & $3.23$ &  & $30.6$ &  & $3.0^{+1.1}_{-1.1}$ &  & $0.31 \textrm{ (fixed)}$ & $0.21^{+0.05}_{-0.03}$ & $3.6^{+1.1}_{-0.6}$ &  & $0.29^{+0.01}_{-0.03}$ & $0.88^{+0.16}_{-0.07}$ & $4.1^{+0.6}_{-0.7}$ &  & $446 / 408$ \\
    \enddata
    \tablecomments{
    This table is published in its entirety in machine-readable format.
    A portion is shown here for guidance regarding its form and content.
    Quoted central values and statistical uncertainties are the posterior medians and 16th--84th percentiles. 
    Values preceded by $<$ are 84th-percentile upper limits.
    Values preceded by $\leq$ indicate that $N_{\rm H}$ was constrained by the imposed upper boundary corresponding to the calculated line-of-sight column density.}
    \tablenotetext{a}{LHB emission measure in units of $10^{-3}\ {\rm cm^{-6}\,pc}$.}
    \tablenotetext{b}{Hydrogen column density in units of $10^{22}\ {\rm cm^{-2}}$.}
    \tablenotetext{c}{Temperature in keV.}
    \tablenotetext{d}{Emission measure of the warm-hot component in units of $10^{-2}\ {\rm cm^{-6}\,pc}$.}
    \tablenotetext{e}{Emission measure of the hot component in units of $10^{-3}\ {\rm cm^{-6}\,pc}$.}
    \end{deluxetable*}
\end{rotatetable*}
\clearpage

\section{Parameter Correlations}
\label{app:parameter_correlations}
\setcounter{figure}{0}

\begin{figure*}[htbp]
    \centering
    \includegraphics[width=0.9\linewidth]{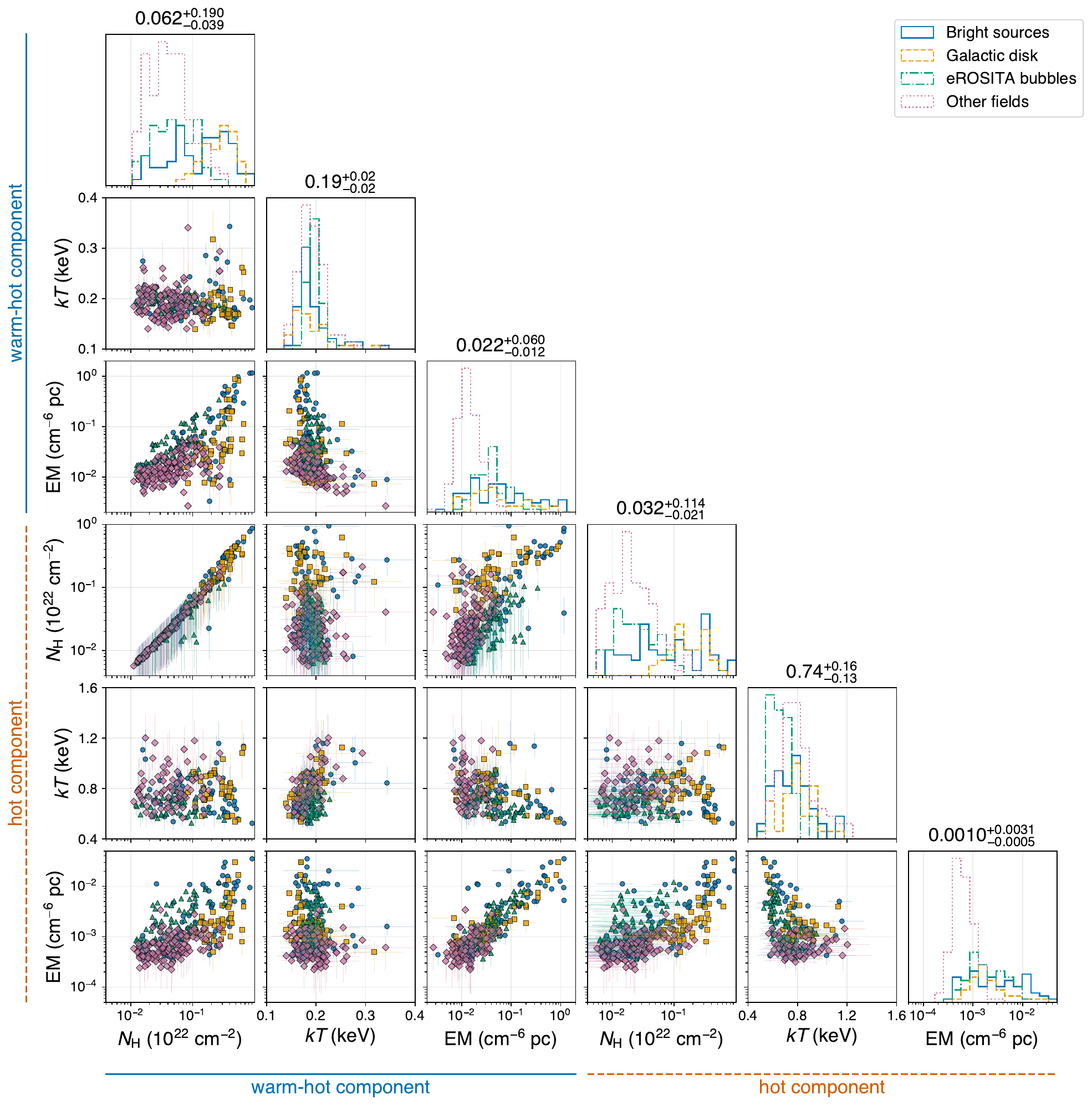}
    \caption{
    Corner plot showing the distributions and pairwise relationships among $N_{\rm H}$, $kT$, and EM for the warm-hot and hot components.
    The colors and marker shapes follow the same spatial classification as in Figure~\ref{fig:corr_em}.
    The diagonal and off-diagonal panels show the parameter distributions and pairwise relationships, respectively.
    For each component, fields with a fixed temperature were excluded from all panels in which that temperature appears on either axis.
    }
    \label{fig:corner_region}
\end{figure*}

Figure~\ref{fig:corner_region} presents a corner plot showing the distributions and pairwise relationships among $N_{\rm H}$, $kT$, and EM for the warm-hot and hot components.
The EM correlation and its dependence on spatial category are discussed in Section~\ref{sec:results} and Figure~\ref{fig:corr_em}, while the other parameter relationships are described below.
For each component, fields with a fixed temperature were excluded from all panels in which that temperature appears on either axis.

The nearly one-to-one relationship between the two $N_{\rm H}$ values mainly reflects the adopted absorption assumption.
The $N_{\rm H}$ value for the warm-hot component was fixed to the calculated value, whereas that for the hot component was constrained not to exceed the same value.
This relationship alone therefore does not allow us to conclude that the two components occupy similar locations along the line of sight.

The temperature correlation coefficient increases from
$\rho=0.36^{+0.04}_{-0.04}$ for the full sample to
$\rho=0.60^{+0.06}_{-0.07}$ for fields in the Other category.
The quoted values are the medians of distributions obtained by sampling the temperatures from their MCMC posteriors, with uncertainties corresponding to the 16th--84th percentiles.
However, this correlation should be interpreted cautiously because of the narrow temperature range of the warm-hot component and the exclusion of fields with fixed temperatures.

\section{All-sky Distribution of the LHB Emission Measure}
\label{app:lhb_distribution}
\setcounter{figure}{0}

\begin{figure}[htbp]
    \centering
    \includegraphics[width=0.98\linewidth]{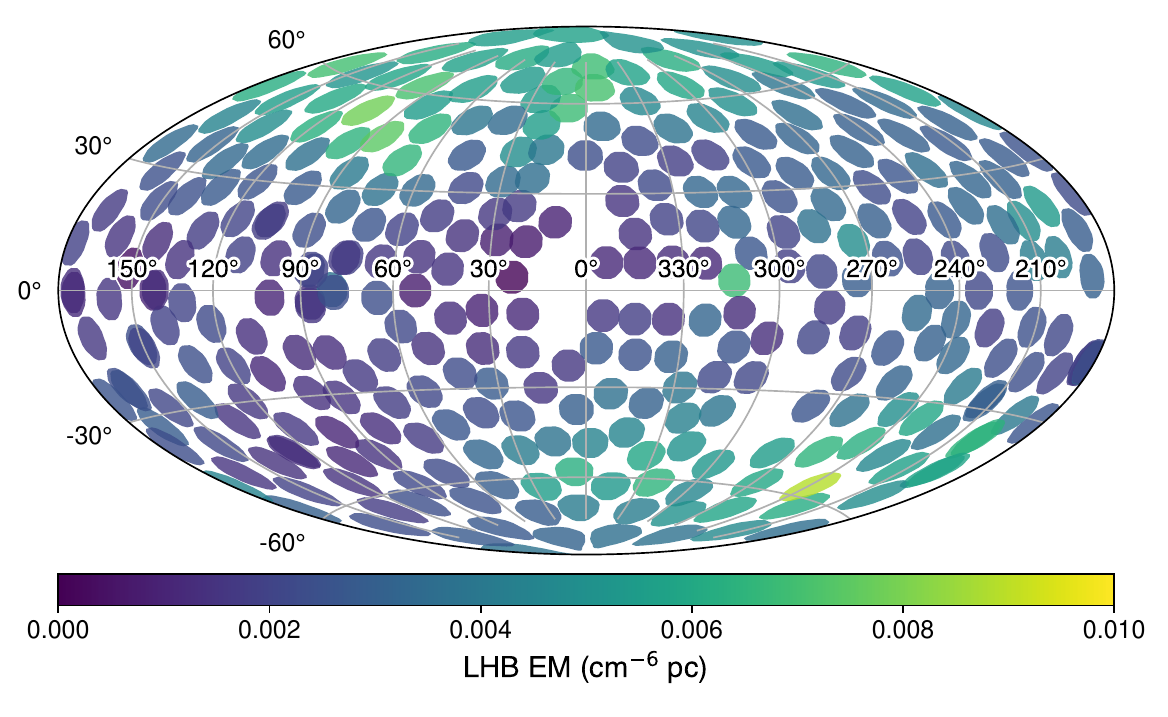}
    \vspace{0.5em}
    \includegraphics[width=0.98\linewidth]{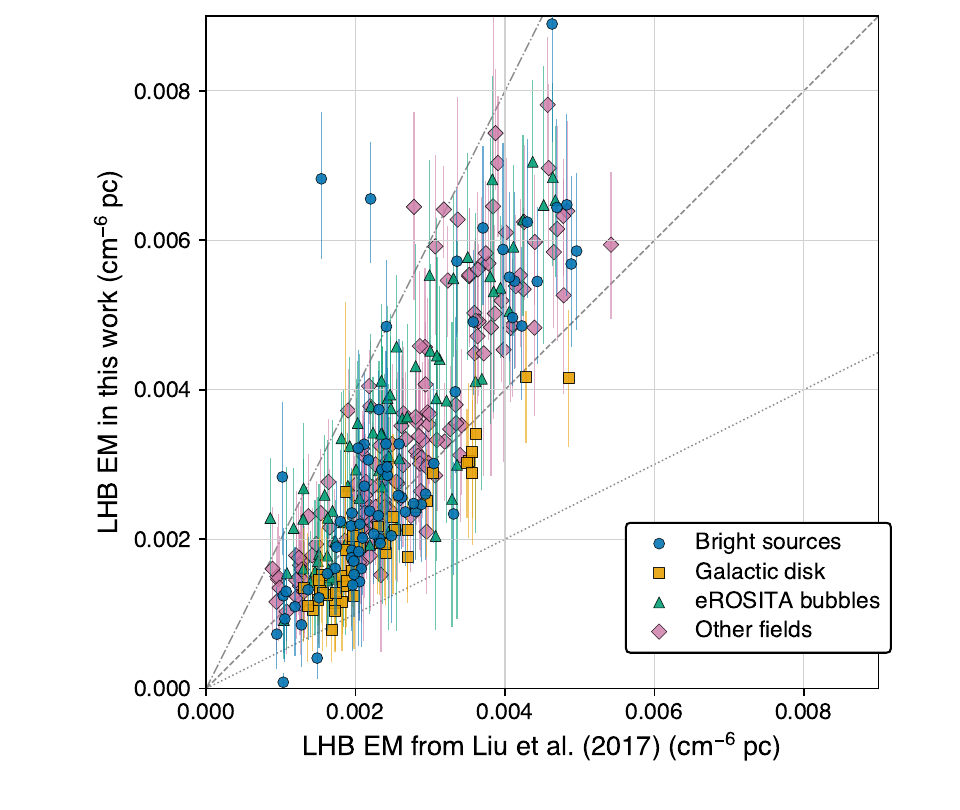}
    \caption{Comparison of the LHB EM distributions.
    The upper panel shows the all-sky distribution of the LHB EM obtained from the spectral fits.
    The lower panel compares the LHB EMs derived from the map of \citet{Liu+2017} with those obtained in this work.
    The dotted lines indicate $y=x$, $y=0.5x$, and $y=2x$.
    The colors and symbols follow the same spatial classification as in Figure~\ref{fig:corr_em}.}
    \label{fig:lhb_appendix}
\end{figure}

Figure~\ref{fig:lhb_appendix} presents the all-sky distribution of the LHB EM obtained from the individual-field spectral fits, together with a field-by-field comparison with the map of \citet{Liu+2017}.
The ratios of the EMs obtained in this work to those reported by \citet{Liu+2017} generally lie between 0.5 and 2.
Within this range, the distribution spans the one-to-one relation and extends toward higher EMs in the present analysis.
The largest deviations occur mainly in fields containing bright X-ray sources and are therefore not representative of the general comparison.
Unlike the band-ratio approach adopted by \citet{Liu+2017}, the present LHB EM distribution was derived by simultaneously modeling the LHB and the other emission components using the \textit{HaloSat} spectra and \textit{ROSAT} R1/R2 constraints.
The resulting map therefore provides an all-sky characterization of the LHB EM based on spectral-component decomposition.

\section{Galactic-longitude Profiles of the Spatial Fitting Results}
\label{app:spatial_profiles}
\setcounter{figure}{0}

\begin{figure*}[htbp]
    \centering
    \includegraphics[width=0.8\linewidth]{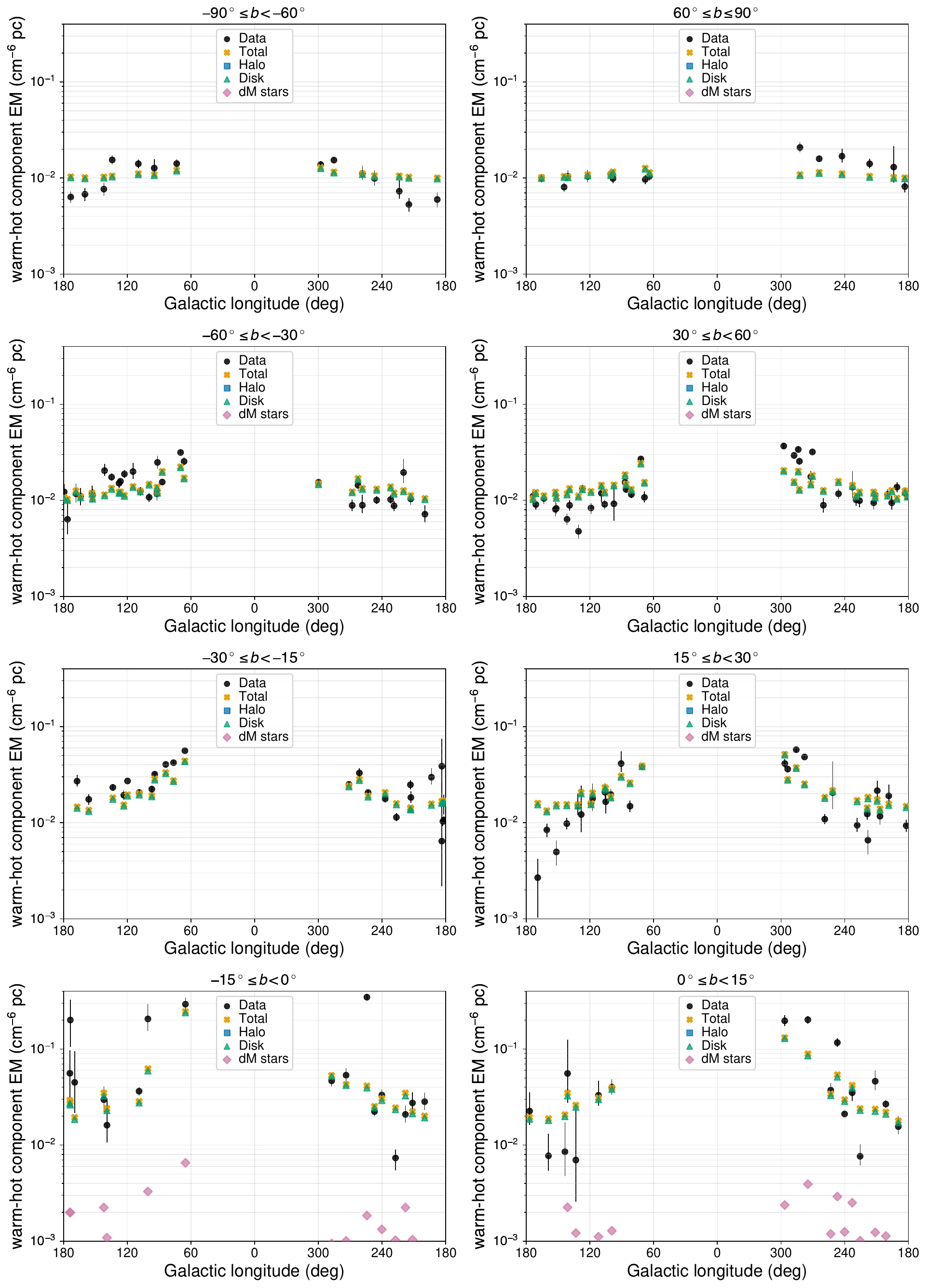}
    \caption{
    Galactic-longitude profiles of the spatial-fitting results for the warm-hot component.
    Black circles with error bars show the observed EMs of the individual fields included in the spatial fit. Orange X-shaped markers show the total model, while blue squares, green triangles, and purple diamonds show the halo-like, disk-like, and dM-star contributions, respectively. 
    The model contributions were evaluated at the Galactic coordinates of each field using the parameters obtained from the global spatial fit.
    The halo-like contribution is constrained only as an upper limit and is included for completeness.
    In some panels, individual model contributions fall outside the displayed EM range and are therefore not visible.
    }
    \label{fig:spatial_profile_02}
\end{figure*}

\begin{figure*}[htbp]
    \centering
    \includegraphics[width=0.8\linewidth]{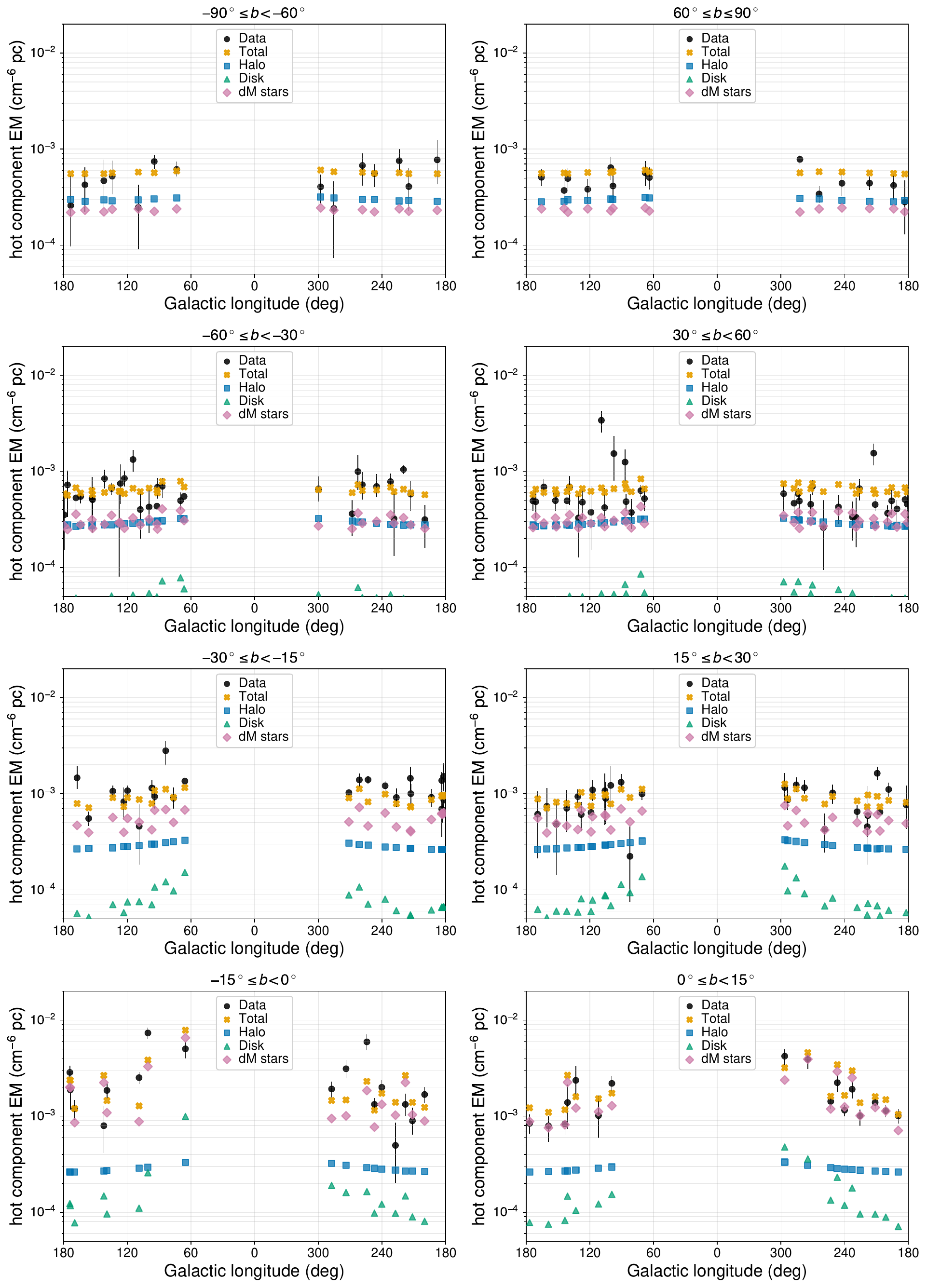}
    \caption{
    Same as Figure~\ref{fig:spatial_profile_02}, but for the hot component.
    The disk-like parameters are not independently constrained because of their degeneracy in the free fit, although a nonzero disk-like contribution remains allowed.
    The displayed contribution is included for completeness.
    }
    \label{fig:spatial_profile_07}
\end{figure*}

Figures~\ref{fig:spatial_profile_02} and \ref{fig:spatial_profile_07} present the spatial-fitting results described in Section~\ref{sec:discussion} as functions of Galactic longitude for the warm-hot and hot components, respectively.
For clarity, the fields are divided into the same eight Galactic-latitude intervals used in the stacked-field analysis.

The observed EMs and the stellar, disk-like, and halo-like model contributions are shown for each field. 
These profiles provide an alternative representation of the global spatial fit shown in Figure~\ref{fig:em_fit}, allowing the longitude dependence of the observed EMs and individual model contributions to be examined within each latitude interval. 
Model components whose parameters are not independently constrained are also shown for completeness.

\clearpage

\end{document}